%% file: main.tex
\documentclass[10pt,twocolumn]{article}
\usepackage{usenix-style}
\usepackage[available,functional,reproduced]{usenixbadges}
\def\noeditingmarks{}

\input{preamble}

\begin{document}

\title{Fault-tolerant and Transactional Stateful Serverless Workflows (extended
  version)\thanks{This is the full version of~\cite{zhang20fault}. 
  This version includes additional details and evaluation results in the 
  appendices.}}

\author{Haoran Zhang, Adney Cardoza$^\dagger$, Peter Baile Chen, Sebastian Angel, and
Vincent Liu\\ {\normalsize \emph{University of Pennsylvania}\quad\quad $^\dagger$\emph{Rutgers
University-Camden}}}
\date{}

\maketitle

\input{abstract}
\input{intro}
\input{bg}
\input{problem}
\input{overview}
\input{workflows}
\input{daal}
\input{api}
\input{collectors}
\input{transactions}
\input{eval}
\input{discussion}
\input{relwork}
\input{conclusion}
\input{acks}

\frenchspacing

\small
\begin{flushleft}
\balance
\setlength{\parskip}{0pt}
\setlength{\itemsep}{0pt}
\bibliographystyle{abbrv}
\bibliography{conferences,paper}
\end{flushleft}

\clearpage

\nobalance
\appendix

\input{appendix}

\label{lastpage}
\end{document}

%% file: preamble.tex
\usepackage{booktabs}  
\usepackage{dcolumn}   
\usepackage{multirow}  
\usepackage{rotating}
\usepackage{xspace}
\usepackage{hyphenat}  
\usepackage{color}
\usepackage[dvipsnames]{xcolor}
\usepackage{enumerate}

\usepackage[square,comma,numbers,sort&compress]{natbib}

\usepackage[normalem]{ulem}      

\usepackage{wrapfig}
\usepackage{textcomp}
\usepackage{lastpage}
\usepackage{tabularx}
\usepackage{pifont}

\usepackage{wrapfig}

\usepackage{balance}

\usepackage{colortbl} 
\usepackage{array}    

\usepackage{dblfloatfix}

\usepackage{listings}
\definecolor{maroon}{cmyk}{0, 0.87, 0.68, 0.32}
\definecolor{halfgray}{gray}{0.55}
\definecolor{ipython_frame}{RGB}{207, 207, 207}
\definecolor{ipython_bg}{RGB}{247, 247, 247}
\definecolor{ipython_red}{RGB}{186, 33, 33}
\definecolor{ipython_green}{RGB}{0, 128, 0}
\definecolor{ipython_cyan}{RGB}{64, 128, 128}
\definecolor{ipython_purple}{RGB}{170, 34, 255}
\definecolor{syellow}{HTML}{B58900}
\definecolor{sorange}{HTML}{CB4B16}
\definecolor{smagenta}{HTML}{D33682}
\definecolor{sviolet}{HTML}{6C71C4}
\definecolor{sblue}{HTML}{268BD2}
\definecolor{scyan}{HTML}{2AA198}
\definecolor{sgreen}{HTML}{859900}

\lstdefinelanguage{iPython}{
    morekeywords={access,and,break,class,continue,def,del,elif,else,except,exec,finally,for,from,global,if,import,in,is,lambda,not,or,pass,print,raise,return,try,while},%
    morekeywords=[2]{abs,all,any,basestring,bin,bool,bytearray,callable,chr,classmethod,cmp,compile,complex,delattr,dict,dir,divmod,does,enumerate,eval,execfile,file,filter,float,format,frozenset,getattr,to,globals,hasattr,hash,help,hex,int,isinstance,issubclass,iter,len,list,locals,long,map,max,memoryview,min,object,oct,open,ord,pow,property,range,raw_input,reduce,reload,repr,reversed,round,set,setattr,slice,sorted,staticmethod,str,sum,super,tuple,type,unichr,unicode,vars,xrange,zip,apply,buffer,coerce,intern,exist,exists},%
    sensitive=true,%
    morecomment=[l]\#,%
    morestring=[b]',%
    morestring=[b]",%
    morestring=[s]{'''}{'''},
    morestring=[s]{"""}{"""},
    morestring=[s]{r'}{'},
    morestring=[s]{r"}{"},%
    morestring=[s]{r'''}{'''},%
    morestring=[s]{r"""}{"""},%
    morestring=[s]{u'}{'},
    morestring=[s]{u"}{"},%
    morestring=[s]{u'''}{'''},%
    morestring=[s]{u"""}{"""},%
    literate=
    {á}{{\'a}}1 {é}{{\'e}}1 {í}{{\'i}}1 {ó}{{\'o}}1 {ú}{{\'u}}1
    {Á}{{\'A}}1 {É}{{\'E}}1 {Í}{{\'I}}1 {Ó}{{\'O}}1 {Ú}{{\'U}}1
    {à}{{\`a}}1 {è}{{\`e}}1 {ì}{{\`i}}1 {ò}{{\`o}}1 {ù}{{\`u}}1
    {À}{{\`A}}1 {È}{{\'E}}1 {Ì}{{\`I}}1 {Ò}{{\`O}}1 {Ù}{{\`U}}1
    {ä}{{\"a}}1 {ë}{{\"e}}1 {ï}{{\"i}}1 {ö}{{\"o}}1 {ü}{{\"u}}1
    {Ä}{{\"A}}1 {Ë}{{\"E}}1 {Ï}{{\"I}}1 {Ö}{{\"O}}1 {Ü}{{\"U}}1
    {â}{{\^a}}1 {ê}{{\^e}}1 {î}{{\^i}}1 {ô}{{\^o}}1 {û}{{\^u}}1
    {Â}{{\^A}}1 {Ê}{{\^E}}1 {Î}{{\^I}}1 {Ô}{{\^O}}1 {Û}{{\^U}}1
    {œ}{{\oe}}1 {Œ}{{\OE}}1 {æ}{{\ae}}1 {Æ}{{\AE}}1 {ß}{{\ss}}1
    {ç}{{\c c}}1 {Ç}{{\c C}}1 {ø}{{\o}}1 {å}{{\r a}}1 {Å}{{\r A}}1
    {€}{{\EUR}}1 {£}{{\pounds}}1
    {^}{{{\color{ipython_purple}\^{}}}}1
    {=}{{{\color{ipython_purple}=}}}1
    {+}{{{\color{ipython_purple}+}}}1
    {*}{{{\color{ipython_purple}$^\ast$}}}1
    {/}{{{\color{ipython_purple}/}}}1
    {+=}{{{+=}}}1
    {-=}{{{-=}}}1
    {*=}{{{$^\ast$=}}}1
    {/=}{{{/=}}}1,
    literate=
    *{-}{{{-}}}1
     {?}{{{\color{ipython_purple}?}}}1,
    identifierstyle=\color{black}\ttfamily,
    commentstyle=\color{ipython_cyan}\ttfamily,
    stringstyle=\color{ipython_red}\ttfamily,
    keepspaces=true,
    showspaces=false,
    showstringspaces=false,
    rulecolor=\color{ipython_frame},
    moredelim=*[is][\color{sblue}\ttfamily]{|}{|},
    moredelim=*[is][\color{syellow}\ttfamily]{$}{$},
    basicstyle=\footnotesize,
    keywordstyle=\color{ipython_green}\ttfamily,
}

\usepackage{algpseudocode}
\algrenewcomment[1]{\hfill// #1}%
\algnotext{EndFunction}
\algnotext{EndFor}
\algnotext{EndIf}

\usepackage{amsmath,amscd}
\usepackage{amssymb}
\usepackage{amsfonts}
\usepackage{amsthm}

\algrenewcommand\algorithmicindent{1em}
\algrenewcommand{\algorithmiccomment}[1]{{\color{CadetBlue}\hfill// #1}}

\usepackage[noeka]{mathrmletter}

\newif\ifextended
\newif\iflongbatching
\newif\ifsubmission
\newif\ifelementary

\ifx\buildextended\undefined
\else
    \extendedtrue
\fi

\ifx\noeditingmarks\undefined
   \newcommand{\pgwrapper}[3]{\begingroup \color{#1} #2: #3 \endgroup}
   \newcommand{\pgwrapperb}[1]{\textbf{#1}}
\else
   \newcommand{\pgwrapperb}[1]{}
   \newcommand{\pgwrapper}[3]{}
\fi

\def\hn{\usefont{OT1}{phv}{mc}{n}\selectfont}

\newcommand{\mpfont}{\hn\scriptsize}

\ifx\noeditingmarks\undefined
    \newcommand{\MPworker}[2]{{\color{#1}\vrule\vrule}{\marginpar{\color{#1}\mpfont #2}}}
\else
    \newcommand{\MPworker}[2]{}
\fi

\ifx\noeditingmarks\undefined
    
\else
    
\fi

\newcommand{\sys}{Beldi\xspace}
\newcommand{\Sys}{\sys}

\theoremstyle{definition}

\makeatletter
\renewcommand*{\@fnsymbol}[1]{\ensuremath{\ifcase#1\or \star\or \dagger\or \ddagger\or
   \mathsection\or \mathparagraph\or \|\or **\or \dagger\dagger
   \or \ddagger\ddagger \else\@ctrerr\fi}}
\makeatother

\newcommand{\C}{\mathcal{C}}

\newcommand{\Z}{\mathbb{Z}}
\newcommand{\Q}{\mathbb{Q}}

\newcommand{\R}{\mathcal{R}}

\makeatletter
\def\imod#1{\allowbreak\mkern10mu({\operator@font mod}\,\,#1)}
\makeatother


\def\compactify{\itemsep=0in \topsep=2pt \parsep=0.00in \partopsep=0pt
\leftmargin=2em}
\let\latexusecounter=\usecounter

  {\begin{list}{\labelitemi}{\itemsep3pt \topsep3pt \parsep0.00in
  \partopsep=3pt \leftmargin1.2em}}%
  {\end{list}}
  {\begin{list}{\labelitemi}{\itemsep1pt \topsep2pt \parsep0.00in
  \partopsep=1pt \leftmargin1.2em}}%
  {\end{list}}
  {\begin{list}{\labelitemi}{\itemsep2pt \topsep2pt \parsep0.00in
  \partopsep=0pt \leftmargin1.2em}}%
  {\end{list}}
  {\begin{list}{\threequartdash}{\itemsep3pt \topsep3pt \parsep0.00in
  \partopsep=3pt \leftmargin1.5em}}%
  {\end{list}}

\newenvironment{myenumerate}
  {\def\usecounter{\compactify\latexusecounter}
   \begin{enumerate}}
  {\end{enumerate}\let\usecounter=\latexusecounter}

\def\compactsortof{\itemsep=0in \topsep=2pt \parsep=0.00in \partopsep=0pt
\leftmargin=1.7em}
\newenvironment{myenumerate2}
  {\def\usecounter{\compactsortof\latexusecounter}
   \begin{enumerate}}
  {\end{enumerate}\let\usecounter=\latexusecounter}

\def\compactsqueeze{\itemsep=0pt \topsep0pt \parsep=0ex \partopsep=0pt
\leftmargin=1.63em}

\ifx\normalpar\undefined
  
\else
  
\fi

\def\discretionaryslash{\discretionary{/}{}{/}}
{\catcode`\/\active
\gdef\URLprepare{\catcode`\/\active\let/\discretionaryslash
        \def~{\char`\~}}}%
\def\URL{\bgroup\URLprepare\realURL}%
\def\realURL#1{\tt #1\egroup}%

\newcommand{\topheading}[1]{\noindent\textbf{#1.}}
\newcommand{\heading}[1]{\vspace{1ex}\noindent\textbf{#1.}}
\newcommand{\weakheading}[1]{\vspace{1ex}\noindent\textit{#1:}}

\usepackage{xstring}
\newcommand{\ensureprefix}[2]{\IfBeginWith{#2}{#1}{\ref{#2}}{\ref{#1#2}}}
\newcommand{\figref}[1]{Figure~\protect\ensureprefix{fig:}{#1}\xspace}

%% file: abstract.tex
\begin{abstract}
This paper introduces \sys, a library and runtime system for writing
  and composing fault-tolerant and transactional stateful serverless functions.
\sys{} runs on existing providers and lets developers
  write complex stateful applications that require fault tolerance and
  transactional semantics without the need to deal with tasks such as 
  load balancing or maintaining virtual machines.
\sys's contributions include extending the log-based fault-tolerant
  approach in Olive (OSDI 2016) with new data structures,
  transaction protocols, function invocations, and garbage collection.
They also include adapting the resulting framework to work over a federated environment 
  where each serverless function has sovereignty over its own data.
We implement three applications on \sys{}, including a movie 
  review service, a travel reservation system, and a social media site.
Our evaluation on 1,000 AWS Lambdas shows that \sys's approach is effective 
  and affordable.
\end{abstract}

%% file: intro.tex
\section{Introduction}\label{s:intro}

Serverless computing is changing the way in which we structure and
  deploy computations in Internet-scale systems.  
Enabled by platforms like AWS Lambda~\cite{lambda}, Azure
  Functions~\cite{azure-functions}, and Google Cloud
  Functions~\cite{google-functions}, programmers can break their 
  application into small functions that providers then
  automatically distribute over their data centers.
When a user issues a request to a serverless-based system, this
  request flows through the corresponding functions to achieve
  the desired end-to-end semantics.
For example, in an e-commerce site, a user's purchase 
  might trigger a product order, a shipping event, a credit card charge, 
  and an inventory update, all of which could be handled 
  by separate serverless functions.

During development, structuring an application as a set of serverless 
  functions brings forth the benefits of microservice architectures: it
  promotes modular design, quick iteration, and code reuse.
During deployment, it frees programmers from the prosaic but 
  difficult tasks associated with provisioning, scaling, and maintaining the 
  underlying computational, storage, and network resources of the system.  
In particular, app developers need not worry about setting up virtual 
  machines or containers, starting or winding down instances to accommodate 
  demand, or routing user requests to the right set of functional 
  units---all of this is automated once an app developer describes
  the connectivity of the units.

A key challenge in increasing the general applicability of serverless computing
  lies in correctly and efficiently composing different functions 
  to obtain nontrivial end-to-end applications.
This is fairly straightforward when functions are stateless, but becomes 
  involved when the functions maintain their own state (e.g., modify 
  a data structure that persists across invocations).
Composing such \emph{stateful serverless functions} (SSFs) 
  requires reasoning about consistency and isolation semantics in the 
  presence of concurrent requests and dealing with component failures.
While these requirements are common in distributed systems and are
  addressed by existing proposals~\cite{corbett12spanner, 
  peng10large, mahmoud13low, zhang15building, mu16consolidating}, SSFs 
  have unique idiosyncrasies that make existing approaches a poor fit.

The first peculiarity is that request routing is stateless. 
Approaches based on state machine replication are hard to implement 
  because a follow-up message might be routed by the
  infrastructure to a different SSF instance 
  from the one that processed a prior message (e.g., an ``accept'' routed 
  differently than its ``prepare''). 
A second characteristic is that SSFs can be independent and have 
  sovereignty over their own data.
For example, different organizations may develop and deploy SSFs, and an 
  application may stitch them together to achieve some end-to-end functionality.
As a result, there is no component in the system that has full visibility 
  (or access) to all the state. 
Lastly, SSF workflows (directed graphs of SSFs) can be complex and include 
  cycles to express recursion and loops over SSFs.
If a developer wishes to define transactions over such workflows (or
  its subgraphs), \emph{all} transactions (including those that will abort)
  must observe consistent state to avoid infinite loops and undefined behavior.
This is a common requirement in transactional memory 
  systems~\cite{spear06conflict, guerraoui08on, mu19deferred, herman16type,
  shamis19fast}, but is seldom needed in distributed transaction protocols

To bring fault-tolerance and transactions to this challenging environment, 
  this paper introduces \emph{\sys}, a library and runtime system for 
  building workflows of SSFs.
\sys{} runs on existing cloud providers without any modification to their 
  infrastructure and without the need for servers.
The SSFs used in \sys{} can come from either the app developer, other 
  developers in the same organization, third-party open-source developers,
  or the cloud providers.
Regardless, \sys{} helps to stitch together these components in 
  a way that insulates the developer from the
  details of concurrency control, fault tolerance, and SSF
  composition. 

A well-known aspect of SSFs is that even though they can persist state,
  this state is usually kept in low-latency NoSQL databases (possibly 
  different for each SSF) such as DynamoDB, Bigtable, and Cosmos DB that 
  are already fault tolerant.
Viewed in this light, SSFs are clients of scalable fault-tolerant
  storage services rather than stateful services themselves.
\sys's goal is therefore to guarantee \emph{exactly-once} semantics to workflows
  in the presence of clients (SSFs) that fail at any point in their execution 
  and to offer \emph{synchronization} primitives (in the form of locks
  and transactions) to prevent concurrent clients from unsafely handling 
  state.

To realize this vision, \sys{} extends Olive~\cite{setty16realizing} 
  and adapts its mechanisms to the SSF setting.
Olive is a recent framework that exposes an elegant abstraction
  based on logging and request re-execution to clients of cloud storage
  systems; operations that use Olive's abstraction enjoy exactly-once semantics.
\sys's extensions include support for operations beyond storage accesses
  such as synchronous and asynchronous invocations (so that SSFs can invoke
  each other), a new data structure for unifying storage of application state 
  and logs, and protocols that operate efficiently on this data 
  structure~(\S\ref{s:daal}).
The purpose of \sys's extensions is to smooth out the differences between 
  Olive's original use case and ours.
As one example, Olive's most critical optimization assumes that clients
  can store a large number of log entries in a database's 
  \emph{atomicity scope} (the scope at which the database can atomically 
  update objects).
However, this assumption does not hold for many databases commonly 
  used by SSFs\@.
In DynamoDB, for example, the atomicity scope is a single row 
  that can store at most 400\,KB of data~\cite{dynamo-limit}---the
  row would be full in less than a minute in our applications.

\sys{} also adapts existing concurrency control and distributed commit
  protocols to support transactions over SSF workflows.
A salient aspect of our setting is that there is no entity that can
  serve as a coordinator: a user issues its request to the first SSF in 
  the workflow, and SSFs interact only with the SSFs in their outgoing 
  edges in the workflow.
Consequently, we design a protocol where SSFs work together (while
  respecting the workflow's underlying communication pattern) to fulfill the 
  duties of the coordinator and collectively decide whether to commit or abort 
  a transaction~(\S\ref{s:transactions}).
 
To showcase the costs and the benefits of \sys, we implement three
  applications as representative case studies: (1) a travel reservation system,
  (2) a social media site, and (3) a movie review service.
These applications are based on DeathStarBench~\cite{gan19open,
  deathstarbench}, which is an open-source benchmark for microservices; we 
  have ported and extended these applications to work without servers 
  using SSFs.
Our evaluation on AWS reveals that, at saturation, \sys's guarantees come at an 
  increase in the median request completion time of 2.4--3.3$\times$, and 
  99th percentile completion time of 1.2--1.8$\times$~(\S\ref{s:eval:apps}).
At low load, the median completion time increase is under 2$\times$.

In summary, \sys{} helps developers build fault-tolerant 
  and transactional applications on top of SSFs at a modest cost.
In doing so, \sys{} simplifies reasoning about compositions of SSFs,
  runs on existing serverless platforms without modifications, and 
  extends an elegant fault-tolerant abstraction.

%% file: bg.tex
\section{Background and Goals}\label{s:bg}

In this section, we describe the basics of serverless computing (sometimes
  known as Function-as-a-Service), the challenge of deploying complex serverless 
  applications that incorporate state, and a list of requirements that
  \sys{} aims to satisfy.

\subsection{Serverless functions}\label{s:bg:serverless}
Serverless computing aims to eliminate the need to manage machines,
  runtimes, and resources (i.e., everything except the app logic).
It provides an abstraction where developers upload a simple function
  (or `lambda') to the cloud provider that is invoked on demand;
  an identifier is provided with which clients and other services
  can invoke the function.

The cloud provider is then responsible for provisioning the VMs or containers,
  deploying the user code, and scaling the allocated resources up and down based on 
  current demand---all of this is transparent to users.
In practice, this means that on every function invocation
  the provider will spawn a new \emph{worker} (VM or container) with the 
  necessary runtime and dispatch the request to this worker (`cold start').
The provider may also use an existing worker, if one is free (`warm start').
Note that while workers can stay warm for a while, running functions
  are limited by a timeout, after which they are killed.
This time limit is configurable
  (up to 15\,min in 1\,s increments on AWS, up to 9\,min in 1\,ms
  increments on Google Cloud, and unbounded time in 1\,s increments on Azure)
  and helps in budgeting and limiting the effect of bugs.

Serverless functions are often used individually, but they can also be
  composed into \emph{workflows}: directed graphs of functions that may 
  contain cycles to express recursion or loops over one or more functions.
Some ways to create workflows include AWS's 
  \emph{step functions}~\cite{step-function} and \emph{driver functions}.
A step function specifies how to stitch together different 
  functions (represented by their identifiers) and their inputs
  and outputs; the step function takes care of all scheduling and data movement,
  and users get an identifier to invoke it.
In contrast, a driver function is a single function specified 
  by the developer that invokes other functions (similar to the 
  main function of a traditional program). 
Control flow can form a graph because functions (including the driver 
  function) can be multi-threaded or perform asynchronous invocations.

\heading{Stateful serverless functions (SSFs)}\label{s:bg:ssm}
Serverless functions were originally designed to be stateless.
As such, state is not guaranteed to persist between function
  invocations---even when writing to a worker's local disk, the function's
  context can be terminated as part of dynamic resource management, or load
  balancing might direct follow-up requests to different or new instances.
Accordingly, a common workaround to persist data is to store it in
  fault-tolerant low-latency NoSQL databases.
For example, AWS Lambdas can persist their state in DynamoDB, 
  Google cloud functions can use Cloud Bigtable, and Azure 
  functions can use Cosmos DB\@.
Through these intermediaries, \emph{stateful serverless functions} (SSFs) can
  save state and expose it to other instances.

Unfortunately, the above approach to state interacts poorly with
  the way that serverless platforms handle failures.
If a function in a workflow crashes or its worker hangs, the
  provider will either (1) do nothing, leaving the workflow incomplete, or
  (2) restart the function on a different worker, potentially incrementing
  a counter twice, popping a queue multiple times, or corrupting database 
  state and violating application semantics.
Indeed, serverless providers currently recommend that developers write SSFs 
  that are idempotent to ensure that re-execution is safe~\cite{google-bestpractice}.
While helpful, these recommendations place the burden entirely on developers.
In contrast, \sys{} simplifies this process so developers need only worry
  about their application logic and not the low-level details of how serverless 
  providers respond to failures.

%% file: problem.tex
\subsection{Requirements and assumptions}\label{s:problem}

We strive to design a framework that helps developers build serverless
  applications that tolerate failures and handle concurrent 
  operations correctly.
Our concrete goals are:

\weakheading{Exactly-once semantics}
\sys{} should guarantee \emph{exactly-once}
  execution semantics in the presence of SSF or worker crash failures.
That is, even if an SSF crashes in the midst of its execution and is restarted
  by the provider an arbitrary number of times, the resulting state 
  must be equivalent to that produced by an execution in which the SSF ran 
  exactly once, from start to finish, without crashing.

\weakheading{SSF data sovereignty}
\sys{} should support SSFs that are developed and managed independently.
For example, multiple instances of an SSF may all access the same database,
  but they might not have access to the databases of other SSFs,
  even those in the same workflow.
Instead, state should only be exposed by choice through an SSF's outputs.
This type of encapsulation is important to support a paradigm in 
  which different developers, organizations, and teams within the same 
  organization are responsible for designing and maintaining their own SSFs\@.
An application developer can then contract with SSF developers (or teams) to
  integrate their SSFs into the application's workflow via the SSF's
  identifier~(\S\ref{s:bg}).
Furthermore, data sovereignty is key to enabling developers to offer 
  proprietary functions-as-a-service to others, and is a best practice in
  microservice architectures~\cite[\S4]{microservice-ebook}.
For example, Microsoft's eShopOnContainers~\cite{microservices-app} serves
  as a blueprint for applying these ideas to real-world applications.

\weakheading{SSF reusability}
\sys{} should allow multiple applications to 
  use the same SSFs in their workflows at the same time. 
This may require each SSF to have different tables or databases to maintain 
  the state of each application separately, though cross-application state
  should also be supported. 

\weakheading{Workflow transactions}
\sys{} should support an optional transactional API that allows an 
  application to specify any subgraph of a workflow that should 
  be processed transactionally with ACID semantics.
We target \emph{opacity}~\cite{guerraoui08on} as the isolation level.
Opacity ensures that (1) the effects of concurrent transactions are 
  equivalent to some serial execution, and (2) every transaction, 
  including those that are aborted, always observes a consistent view of the database.
We discuss why these requirements are important in SSFs in
  Section~\ref{s:transactions:tx}.

\weakheading{Deployable today}
\sys{} should work on existing serverless platforms without any modifications 
  on their end.
This allows developers to use \sys{} on any provider of their choosing (or 
  even a multi-provider setup), and lowers the barrier to switch providers.
Additionally, developers should not need to run any servers in 
  order to use \sys.
After all, a big appeal of serverless is that it frees developers
  from such burdens.

\paragraph{Assumptions.}
To achieve these goals, \sys{} makes some assumptions about
  the storage provided to SSFs: that it supports strong consistency, 
  tolerates faults, supports atomic updates on some atomicity scope (e.g., 
  row, partition), and has a scan operation with the ability to filter 
  results and create projections.
These assumptions hold for the NoSQL databases commonly used by SSFs: 
  Amazon's DynamoDB, Azure's Cosmos~DB, and Google's Bigtable.

%% file: overview.tex
\section{Design Overview}\label{s:overview}\label{s:design}

\sys{} consists of a library that developers use to write their 
  SSFs and a serverless-based runtime system to support them.
\sys's approach to handling SSF failures is based on an idea most
  recently explored by Olive~\cite{setty16realizing} and 
  inspired by decades of work on log-based fault 
  tolerance~\cite{gray78notes, mohan92aries}.
Specifically, \sys{} executes SSF operations while atomically logging these 
  actions and periodically re-executes SSFs that have not yet finished.
The logs prevent duplicating work that has already been done,
  guaranteeing \emph{at-most-once} execution semantics, while the 
  re-execution ensures \emph{at-least-once} semantics.
 
\input{fig-arch}

Figure~\ref{fig:arch} depicts \sys's high-level architecture.
  \sys{} consists of four components: (1) the \sys{} library, which exposes 
  APIs for invocations, database reads/writes, and transactions; (2) a 
  set of database tables that store the SSF's state, as well as logs 
  of reads, writes, and invocations; (3) an \emph{intent collector}, which is a 
  serverless function that restarts any instances of the 
  corresponding SSF that have stalled or crashed; and (4) a 
  \emph{garbage collector}, which is a serverless function that keeps the 
  logs from growing unboundedly.

To ensure data sovereignty (\S\ref{s:problem}), the runtimes and logs of 
  different SSFs are independently managed and stored; however, 
  all instances of \textit{related} SSFs may share the same
  \sys{} infrastructure.
An app developer composes multiple SSFs into a workflow by
  chaining them together using a driver function, a step
  function, or a combination of the two.
In the following sections we expand on each of these components.

\subsection{Initial inspiration: Olive}\label{s:olive}
Olive~\cite{setty16realizing} guarantees exactly-once execution semantics 
  for clients that may fail while interacting with a fault-tolerant storage
  server.
This is a similar objective as ours, though our setting makes applying
  Olive's ideas nontrivial.
An intent in Olive is an arbitrary code snippet that the client intends to
  execute with exactly-once semantics.
Each intent is assigned a unique identifier (\emph{intent id}), which Olive 
  uses as the primary key to save its progress.
A client in Olive enjoys \emph{at-most-once} semantics by checking the intent's 
  progress and skipping completed operations during re-execution.
Intents consist of local and external operations.
For example, incrementing a local variable is a local operation, whereas
  reading a value from storage is an external operation.
Each external operation in the intent is assigned a monotonically increasing
  \emph{step number}, starting at 0, that uniquely identifies it.

There are two key requirements for intents.
First, intents must be deterministic; developers can make non-deterministic 
  operations (e.g., a call to a random number generator) deterministic by 
  logging their results and replaying the same values in the event of a 
  re-execution.
Second, intents must be guaranteed to always complete in the absence 
  of failures (e.g., they must be free from bugs such as deadlock or infinite loops).

After an intent has been successfully logged, the client in Olive executes
  the intent's code normally until it reaches an external operation 
  (e.g., reading or writing to the database).
Then, the client:
  (1) determines the operation's step number;
  (2) performs the operation (e.g., writes to the database);
  (3) logs the intent id, step number, and the operation's return 
      value (if any) into a separate database table called the 
      \emph{operation log}.
When the client completes all operations, it marks the intent as `done' 
  in the intent table.

To ensure at-most-once execution semantics, the client in Olive must 
  perform actions (2) and (3) above atomically. 
This ensures that if Olive re-executes an intent, there will be a record
  in the operation log showing that a particular step has already been
  completed and should not be re-executed.
Instead, the entity re-executing the intent should resume execution from the
  last completed step, using logged return values from previous steps as needed.
To make these two actions atomic, Olive introduces a technique called
  \emph{Distributed Atomic Affinity Logging} (DAAL), which collocates log
  entries for an item in the same \emph{atomicity scope} (the scope at which
  the database supports atomic operations) with the item’s data.
For example, in a storage system where operations are atomic at the row 
  level, Olive would store the item's value and its log entries in different
  columns of the same row.

\heading{Intent collector}
To guarantee \emph{at-least-once} semantics, Olive must ensure that some
  entity finishes the intent if the client crashes.
This is the job of the \emph{intent collector} (IC), which is a background
  process that periodically scans the intent table and completes 
  unfinished intents by running their code.
Before the IC executes an external operation, it consults the operation log 
  table with the operation's step number to see if the operation has 
  already been done and to retrieve any return value; regular clients also 
  perform this check between actions (1) and (2).
If the operation has not been done, the IC atomically executes
  the operation and logs the result to the operation log table.
Even if multiple IC instances execute concurrently, or if the 
  IC starts executing the intent of a client that has not crashed,
  this is safe because of Olive’s assurance of at-most-once semantics.

\heading{\sys{} vs. Olive}
\sys{} is inspired by Olive's high-level approach but makes key changes and 
  introduces new data structures and tables, support for invocations 
  so that SSFs can call each other (Olive only supports storage operations), 
  and garbage collection mechanisms to keep overheads low.

An important difference between the two is the definition of an `intent.'
In Olive, intents are code snippets---logged by the client---and all intents are
  logged in the same intent table.
In \sys{}, the \emph{client} (which is the SSF) is the code snippet.
As a result, an intent in \sys{} is not code but rather the parameters that
  identify a particular running instance of the SSF\@: its
  inputs, start time, whether it was launched asynchronously, etc.
Accordingly, \sys{} uses the term `instance id' instead of `intent id'
  to capture this distinction.

Another critical difference is that, as shown in Figure~\ref{fig:arch}, each 
SSF in \sys{} is backed by a different database and \sys{} runtime to ensure data 
sovereignty, though different SSFs developed by the same engineering 
team may reuse these components if desired.
We will expand on these details in the following sections, but we
  begin by introducing \sys's API\@.

%% file: fig-arch.tex
\begin{figure}
  \centering
  \includegraphics[width=0.48\textwidth]{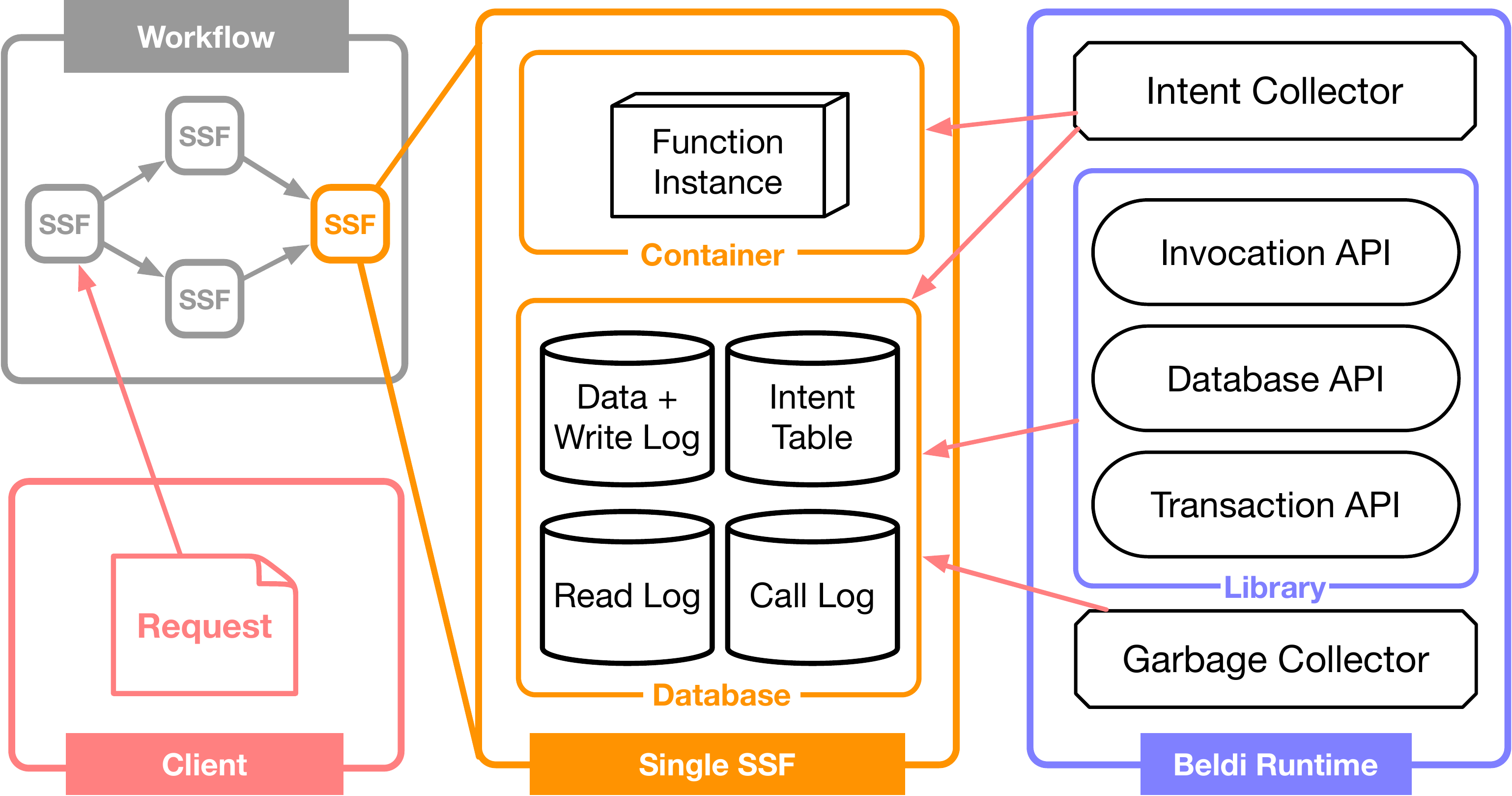}
\caption{\sys's architecture. Developers write SSFs as they do today, 
    but use the \sys{} API for transactions and externally visible operations.
At runtime, operations for each SSF are logged to a database, which, when 
  combined with a per-SSF intent and garbage 
  collector, guarantees exactly-once semantics.
}%
\label{fig:arch}
\end{figure}

%% file: workflows.tex
\subsection{\sys's API}\label{s:api}

\input{table-api}

\sys{} exposes the API in Figure~\ref{fig:api}, which
  includes key-value operations such 
  as \texttt{read}, \texttt{write}, and \texttt{condWrite} (a write that 
  succeeds only if the provided condition evaluates to true), 
  and functions to invoke other SSFs (\texttt{syncInvoke} and 
  \texttt{asyncInvoke}).
These operations are meant as drop-in replacements for the existing
  interface used by SSFs\@.
Furthermore, \sys{} supports the ability to \texttt{begin} and 
  \texttt{end} transactions; operations between these calls enjoy ACID 
  semantics.

\sys's API hides from developers all of the complexity of logging, 
  replaying, and concurrency control protocols that take place 
  under the hood to guarantee exactly-once semantics and support transactions.
For example, an SSF using \sys's API automatically determines
  (from the input, environment, and global variables) the SSF's instance id, 
  step number, and whether it is part of a transaction.
\sys{} takes actions before and after the main body of the SSF as well as around any \sys API operations.

\subsection{\sys's runtime infrastructure}\label{s:workflow}
Developers write SSF code as they do today, but link \sys's library
  and use its API\@.
The rest of \sys's mechanisms happen behind the scenes.
  
\heading{Intent table}
\sys{} associates with every SSF invocation an 
  \emph{instance id}, which uniquely identifies an intent to execute a given
  SSF\@.
For the first SSF in a workflow, the instance id is the UUID assigned by the
  serverless platform to the initial request.
For example, in AWS this UUID is called the `request id,' in GCP it is called
  the `event id,' and in Azure it is the `invocation id.'
For subsequent SSFs in the workflow, each caller in the graph will generate
  a new UUID to be used by the callee as its instance id.
A new id is generated even if the SSF has been invoked earlier in the workflow
  or if the callee is another instance of the caller SSF (in the case of
  recursive functions).
Thus, every SSF instance will have a distinct instance id, even if the instances
  are of the same SSF and in the same workflow.

\sys{} keeps an intent table that contains the instance id, arguments, completion
  status, and other information listed in Figure~\ref{fig:logs} for every SSF
  instance that users and other SSFs intend to execute.
It does this by modifying SSFs to ensure that the first operation 
  is to check the intent table to see if their instance id 
  is already present and, if not, to log a new entry.
\sys{} performs a similar modification to set the intent as `done' at the 
  end of the SSF execution.

\input{table-logs}

\heading{Operation logs}
In addition to the intent table, \sys{} maintains three logs 
  for each SSF\@: a \emph{read log}, \emph{write log}, and \emph{invoke log}.
Their schema is also in Figure~\ref{fig:logs}.
For each operation, the key into the log is the combination of the executing
  SSF's instance id and the step number, which (like in Olive) is a counter 
  that identifies each unique external operation.
Each read operation adds the value read from the database into the read log.
Writes, meanwhile, write to the write log with a boolean flag
  that states whether the write operation took effect.
Regular writes always set this flag to true, while conditional writes 
  set it to the outcome of the condition at the time of the write. 
The actual data being written is stored in a data table, although
  in Section~\ref{s:daal} we discuss a data structure that generalizes
  Olive's DAAL and collocates the write log in the same table as the data 
  to avoid cross-table transactions.
The invoke log is new to \sys{} and ensures at-most-once semantics for calls 
  to other SSFs; we describe it in Section~\ref{s:invoke}.

\heading{Intent and Garbage Collectors}
For each SSF, \sys{} introduces a pair of serverless functions that are
  triggered periodically by a timer.
The first function acts as the SSF's \emph{intent collector} (IC).
The IC scans the SSF's intent table to discover instances of the SSF
  that have not yet finished (lack the `done' flag).
The IC restarts each unfinished SSF by re-executing it with the
  original instance id and arguments.
Note that it is safe for the IC to restart an SSF instance even if the 
  original instance is still running and has not crashed, owing to \sys's use 
  of logs to guarantee at-most-once semantics for each step of the SSF\@.
We implement two natural optimizations for the IC\@.
First, the IC restarts instances only after some amount of time
  has passed since the last time they were launched to avoid spawning
  too many duplicate instances in cases where the IC runs very frequently.
Second, the IC speeds up the process of finding unfinished instances among
  all instance ids in the intent table by maintaining a secondary index.

The second function acts as a \emph{Garbage Collector} (GC) for completed intents,
  taking care to ensure safety in the presence of concurrent SSF instances,
  IC instances, and even GC instances.
This component is described in Section~\ref{s:garbage}.

%% file: table-api.tex
\begin{figure}
{\small
\begin{tabularx}{\columnwidth}{l  X}
\textbf{\sys Library Function} & \textbf{Description} \\
\toprule
\texttt{read(k)} $\to$ \texttt{v} & Read operation\\
\texttt{write(k, v)} & Write operation\\
\texttt{condWrite(k, v, c)} $\to$ T/F & Write if \texttt{c} is true\\
\texttt{syncInvoke(s, params)} $\to$ \texttt{v} & Calls \texttt{s} 
and waits for answer\\
\texttt{asyncInvoke(s, params)} & Calls \texttt{s} without waiting \\
\midrule
\texttt{lock()} & Acquire a lock\\
\texttt{unlock()} & Release a lock\\
\texttt{begin\_tx()} & Begin a transaction\\
\texttt{end\_tx()} & End a transaction\\
\bottomrule
\end{tabularx}
}
\caption{\Sys's API for SSFs, which includes all of \sys's primitives and
  its transactional support~(\S\ref{s:transactions}).}
\label{fig:api}
\end{figure}

%% file: table-logs.tex
\begin{figure}
{\small
\begin{tabularx}{\columnwidth}{l l l}
\textbf{Log} & \textbf{Key} & \textbf{Value} \\
\toprule
\texttt{intent} & instance id & done, async, args, ret, ts \\  
\texttt{read} & instance id, step number & value \\
\texttt{write} & instance id, step number & true / false \\
\texttt{invoke} & instance id, step number & instance id of callee, result \\
\bottomrule
\end{tabularx}
}
\caption{\Sys{} maintains four logs for each SSF\@. 
The intent table keeps track of an instance's completion status, 
  arguments, return value, type of invocation, and
  timestamp assigned by its garbage collector (ts).
The read log stores the value read.
The write log stores true for writes, or the
  condition evaluation for a conditional write.
The invoke log stores the instance id of the callee 
  and its result.}%
\label{fig:logs}
\end{figure}

%% file: daal.tex
\section{Executing and Logging Operations in \sys}\label{s:daal}

As we mention in Section~\ref{s:olive}, guaranteeing
  exactly-once semantics requires atomically logging and
  executing operations.
This section discusses how \sys{} achieves this.

\subsection{Linked DAAL}\label{s:linked-daal}
The logging approach taken by Olive (\S\ref{s:olive}) requires an 
  \emph{atomicity scope} with high storage capacity, as otherwise few 
  log entries can be added.
In the context of Cosmos DB (the successor of the database
  used by Olive), the atomicity scope is a database partition, and
  the atomic operation is a transactional batch update.
Olive's DAAL is a good fit for Cosmos DB because partitions 
  can hold up to 20~GB of data~\cite{cosmos-quotas}, 
  which is enough to collocate a data item and a large number of log entries.
However, other databases adopt designs with more limited atomicity scopes.
For example, the atomicity scope of DynamoDB and Bigtable is one 
  row, which can hold up to 400~KB~\cite{dynamo-limit} and 
  256~MB~\cite{bigtable-quota}, respectively; the recommended limits 
  are much lower.
If we were to use Olive's DAAL with DynamoDB, an SSF could only perform 
  hundreds of writes to a given key before filling up the row.
At such point, Olive would be unable to make further progress until 
  the logs are pruned.
This is hard to do in our setting: reaching a state of 
  quiescence where it is safe to garbage collect logs is challenging 
  since existing platforms expose no mechanism to kill or pause 
  SSFs~(\S\ref{s:garbage}).

To support all common databases, \sys{} introduces a new data 
  structure called the \emph{linked DAAL} that allows logs to 
  exist on multiple rows (or atomicity scopes), with 
  new rows being added as needed.
There are three reasons why this simple data structure is 
  interesting for our purposes: 
  (1) linked DAALs continue to avoid the overheads of 
      cross-table transactions and work on databases that do not
      support such transactions;
  (2) linked DAALs are a type of non-blocking linked 
      list~\cite{valois95lock, harris01pragmatic, zhang13practical}, 
      allowing multiple SSFs to access them concurrently with 
      the operations supported by the atomicity scope (e.g., atomic 
      partial updates);
  (3) even with frequent accesses, our garbage collection protocol
      can ensure that the length of the list for each item is kept
      consistently small~(\S\ref{s:garbage}).

\input{fig-linkeddaal}

\heading{Structure}
Figure~\ref{fig:linked-daal} gives an example of a linked DAAL for an 
  item with two rows of logs.
Every row stores the item's key, value, owner of the lock (used for 
  transactions in Section~\ref{s:transactions}), the log of writes,
  and metadata needed to traverse the linked DAAL and perform garbage collection.
The first row is the `head,' which has a special \texttt{RowId} and is 
  never garbage collected.
The primary key for rows is \texttt{RowId} + \texttt{Key}, the hash
  key is \texttt{Key}, and the sort key is \texttt{RowId}.
When a row is full and the SSF issues a write operation, a new
  row is appended with the updated value and a log entry describing
  the write; the previous row's value and logs are not modified 
  once filled.
Thus, the tail always has the most recent value.

\heading{Traversal}
Most operations in \sys{} require traversal to the tail of the list.
The simplest way to accomplish this is to start at the designated
  head row and iteratively issue read requests for 
  each \texttt{NextRow} until the field is empty.
While this procedure will eventually reach the tail, the number
  of database operations grows with the length of the list.
Garbage collection can control this length, but \sys{} applies an additional
  optimization that leverages the scan and projection operations available
  in the three NoSQL databases that we surveyed.
Specifically, \sys{} issues a single scan operation to the database
  that returns every row containing a target \texttt{Key}.
On its own, the scan operation returns all contents of each row (including the values, write logs, etc.).
To reduce this overhead, \sys{} applies a projection that filters out all columns except for 
  \texttt{RowId} and \texttt{NextRow}.
This combination of scan and projection allows \sys{} to download only
  256 bits per row of the linked DAAL\@.
From these rows, \sys{} constructs a skeleton version of the linked
  DAAL locally, which it can quickly traverse to find the \texttt{RowId} of the
  tail.

We note that the individual reads in a scan are not executed atomically.
For example, \sys{} might see a row with no \texttt{NextRow}, and also receive 
  a row that was subsequently appended to it.
This operation might even retrieve rows that are orphaned from a 
  failed append operation.
Regardless, when these databases are configured to be 
  linearizable~\cite{cosmos-consistency,dynamo-consistency,bigtable-consistency},
  the set of rows traversed from the head to the first instance of an empty
  \texttt{NextRow} form a consistent snapshot of the linked DAAL---any write
  that completes strictly before the scan begins will be reflected in the
  constructed local linked DAAL\@.

While the linked DAAL is structurally simple, operating on it requires
  care.
The following sections detail how \sys's API functions read and modify
  the linked DAAL\@.

%% file: fig-linkeddaal.tex
\begin{figure}
	\centering
  \includegraphics[width=0.85\columnwidth]{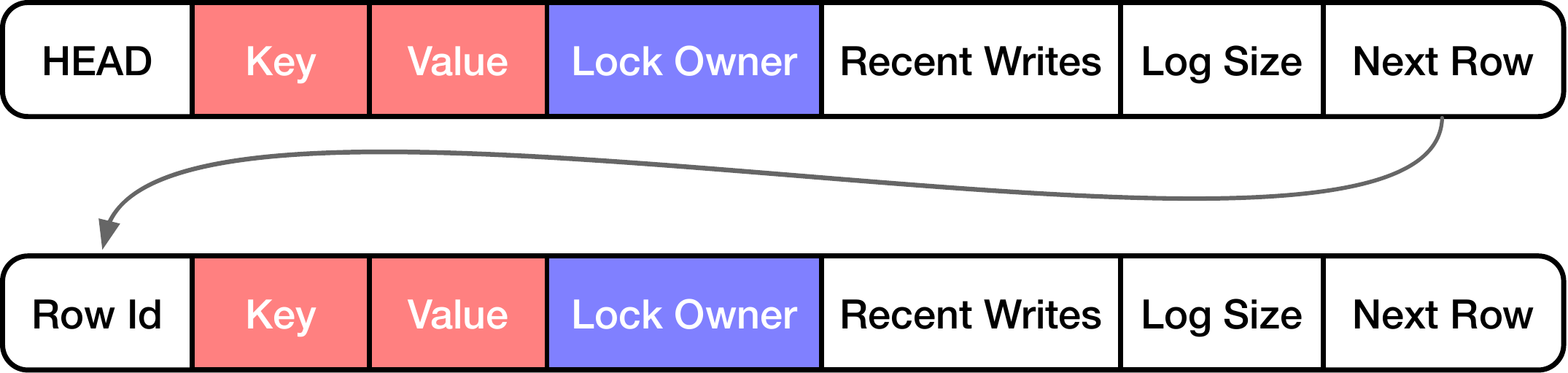}
  \caption{Linked DAAL for a single item. Each row 
    contains the item's key, previous values (except the last row which
    contains the current value), lock information (used for transactions), 
    a log of recent writes, and 
    information for traversal and garbage collection.
  }%
  \label{fig:linked-daal}
\end{figure}

%% file: api.tex
\subsection{Read}%
\label{s:read}

\input{alg-read}

We begin by discussing \sys's \texttt{read} operation.
While \texttt{read} has no externally visible effects on its own, 
  the potential use of its non-deterministic results in a subsequent 
  external operation means that \sys{} must record the result of 
  every \texttt{read} in a dedicated \emph{ReadLog}.
Unlike write operations, however, the read from the database and the 
  log to the ReadLog need not happen atomically---if the SSF crashes 
  before logging the outcome, it is fine to fetch a 
  fresh value as the previous result did not have any externally visible 
  effect.

Figure~\ref{fig:alg-read} shows the pseudocode of the \texttt{read} API 
  function, which involves two steps: (1) read the most recent 
  value of the key from the tail of the linked DAAL, and (2) log the 
  result to the ReadLog if it has not yet been completed.
For the first step, \sys{} retrieves the tail as described in 
  Section~\ref{s:linked-daal}.
For the second step, \sys{} uses an atomic conditional update to efficiently
  log the operation without overwriting a previously executed read.
If it encounters a conflict during the update, it returns the
  previous result from the ReadLog.

\subsection{Write}%
\label{s:write}

\input{alg-write}

A write is more complex as the update and logging must be
  done atomically---within the same atomicity scope---and 
  \sys{} needs to handle cases where other SSFs are accessing and appending to 
  the linked DAAL concurrently.
At a high level, the write operation must find the tail of 
  the linked DAAL, check if the write has been previously executed, 
  log/update the tail if it has not, and extend the 
  linked DAAL if the current tail is full.
Like \texttt{read}, \sys{} can use scan and projection to assemble a 
  minimal local version of the linked DAAL\@.
Unlike \texttt{read}, \sys{} cannot skip directly to the tail;
  instead, \sys{} must check that none of the scanned rows contains a 
  record of the current operation.
Furthermore, once \sys{} has a candidate for the tail, \sys{} needs to 
  update its value and add an entry to its log atomically.
For a given tail candidate there are exactly four possible scenarios:

\input{fig-writestate}

\begin{myenumerate}
\item[A.] The operation has already been executed and 
  the [instance ID, step number] tuple can be found in the current row.  
  \sys{} can return immediately in this case.
\item[B.] The operation is not in the log and there is still space. 
  This indicates that \sys{} is at the tail,
  the operation has never been executed previously, and there is 
  room in the current row to execute/log the write.
\item[C.] The operation is not in the log, but the log is full and
  there is a pointer to the next row. \sys{} should follow the 
  provided pointer toward the tail.
\item[D.] The operation is not in the log and the log is full, but 
  there is no next row.  \sys{} should append a new row and advance to that 
  new row.
\end{myenumerate}

We formulate a lock-free algorithm to handle all the cases above
  by examining the transitions induced by concurrent SSF accesses.
For example, if \sys{} is in case B, where the operation is not in 
  any log and there is still space to execute it in the current row, 
  a concurrent SSF can, without warning, execute the current 
  operation ($\rightarrow$A) or fill the remaining space in the log
  ($\rightarrow$C/D).
The reverse is not true: once there is a \texttt{NextRow} pointer, the 
  linked DAAL will never revert to having extra space for logs.
The cases and their transitions are summarized in \figref{fig:write_states},
  where $N$ is the maximum number of log entries that can fit in a row when 
  accounting for the size of the key, value, and other metadata.
The exception is garbage collection (not covered in \figref{fig:write_states}),
  whose operation and correctness we describe in Section~\ref{s:garbage}.
An arrow in \figref{write_states_transitions} indicates a possible 
  effect of concurrent SSF instances.

To safely identify the state of a row, \sys{} checks for each case
  starting at the node(s) in the transition graph without incoming edges.
In this case there is only one such node (B), so \sys{} performs a
  conditional write with the condition given in case B of
  \figref{write_states_cases} (i.e., that the logKey is not in
  the logs, that the logSize is less than N, and that there is no
  nextRow).
If the conditional check fails, the state of the row will not revert back
  to case B later because B has no incoming edges. 
Therefore, it is safe to remove B from the transition graph and check 
  the remaining cases.
\sys{} repeats the above process with cases A and D (in any order) because 
  they they have no incoming edges in the remaining graph.
Finally, if all prior conditions fail, the row is in case C.

\subsection{Conditional write}%
\label{s:cond-write}

\sys{} also provides support for conditional writes, which only 
  execute if a user-defined condition is true at the time of the write.
The initial scan and subsequent scenarios are similar to the scenarios for
  unconditional writes.
The only exception is the case where the operation has not 
  previously executed and the current row still has remaining space 
  in the log (i.e., case B from Section~\ref{s:write}).
We split this case into two: in B$_1$, the condition is true, 
  and in B$_2$, the condition is false.
\sys{} handles these cases by first checking B$_1$ and B$_2$ with conditional 
  writes before covering the other states exactly as in the unconditional-write case.
We give a detailed description in Appendix~\ref{s:appendix-impl}.

\subsection{Invocation of SSFs and local functions}\label{s:invoke}

Finally, \sys{} supports three types of function invocations:
  synchronous calls (\texttt{syncInvoke}), which block and return a value; 
  asynchronous calls (\texttt{asyncInvoke}), which return
  immediately; and calls to functions that do not use \sys's 
  API (e.g., legacy libraries or legacy SSFs).
In the first two cases, \sys{} guarantees exactly-once semantics.
In the last, it only guarantees that the operation is performed at least once.

\input{alg-syncinvoke}

\figref{fig:alg-synccall} shows pseudocode for synchronous SSF invocations.
As mentioned in Section~\ref{s:workflow}, to help SSFs that are being invoked
  (``callees'') differentiate between re-executions and new executions,
  \sys{} passes an instance id to the callee (the ``callee id'') along with the
  parameters of the call.
In the first invocation, the callee id is generated using \texttt{UUID()}; for
  re-executions, it retrieves the id from the invoke log.
If there is already an entry in the invoke log for this caller id and step number,
  there are two cases: (1) a result 
  is already present, in which case the caller reuses that result; 
  or (2) the entry is present but there is no result, in 
  which case the caller re-invokes the callee with the 
  existing callee id.

\input{fig-callback}

\heading{Callbacks}
Note that \texttt{syncInvoke} (\figref{fig:alg-synccall}) does not
  log the result of the actual call or otherwise mark the call as complete.
To see why this is important, consider the example trace in \figref{fig:callback},
  which shows the result of a failure of the callee (SSF2) after it marks itself as
  done in the intent table but before it returns the result to the caller (SSF1).
Suppose that there is no callback, i.e., that SSF2 logs itself as complete
  immediately after completing execution.
\sys's federated setup means that each SSF has a garbage collector running at its 
  own pace.
If SSF2 were to fail after logging itself as done, it is, therefore, possible
  that SSF2's GC will garbage collect the intent before SSF1 gets any value.
Later, when SSF1's IC re-executes the unfinished SSF1 instance, the caller
  will see the lack of result in the invoke log, re-invoke SSF2 (with the 
  existing callee id), and SSF2 will mistakenly perform the operation again.
In some ways, this is similar to why \texttt{write} operations 
  in \sys{} must be atomically logged and executed~(\S\ref{s:olive}).
Unfortunately, there are no mechanisms for atomically logging into a database
  and executing other SSFs.

We address this issue by decomposing an invocation into two steps: 
  (1) the invocation itself, performed by the caller; and (2) the recording of
  results, done via a second, automatic invocation by the callee to \emph{some}
  instance of the caller.
We emphasize `some' and `original' because request routing in serverless
  is stateless: if SSF1 invokes SSF2, and SSF2 then invokes SSF1, 
  the two SSF1 instances could be different~(\S\ref{s:bg:serverless}).
We call this automatic invocation a \emph{callback}.
When the second instance of the caller receives the callback, it logs the
  provided result in its invoke log and returns.
At this point, it is safe for the callee to mark its intent as 
  done since it knows the caller's invoke log already contains the result.
Note that callbacks only require at-least-once semantics, so there
  is no need for additional logging of the callback invocation.

\figref{fig:callback} illustrates the idea of \sys's callback mechanism.
The callback ensures that the result of SSF2 is properly received by SSF1.
As such, we note that SSF2's response to SSF1 is merely an optimization
and not necessary for correctness.
We also note that if SSF2 fails after a successful callback but before logging the completion of 
the intent, it may result in a case where SSF1 completes, gets garbage collected, and then a 
re-execution of SSF2 invokes a spurious callback.
SSF1 can detect and ignore this case when a callback occurs for an invoke that does 
not exist.

\heading{Asynchronous invocations}
This procedure is similar to that of synchronous invocations, but with the two
  steps flipped on the callee.
The caller first makes a \texttt{rawSyncInvoke} call to the callee, but rather
  than execute the function, the callee (observing an `async' flag) simply
  registers the intent in its intent table, issues a callback, 
  and then immediately returns to the caller.
In the second step, \sys{} performs the actual asynchronous invocation of SSF2's logic.
We describe this operation in detail in Appendix~\ref{s:appendix-impl}.

%% file: alg-read.tex
\begin{figure}
\hrule
\begin{lstlisting}[language=iPython]
 def |read|(table, key):
     linkedDAAL = |rawScan|(table,
         cond: "Key is {key}",
         project: ["RowId", "NextRow"])
     tail = |getTail|(linkedDAAL)
     val = |rawRead|(table, tail)
     logKey = [$ID$, $STEP$]
     $STEP$ = $STEP$ + 1
     ok = |rawCondWrite|($ReadLog$, logKey,
         cond: "{logKey} does not exist"
         update: "Value = {val}")
     if ok:
         return val
     else:
         return |rawRead|($ReadLog$, logKey)
\end{lstlisting}
\hrule\vspace{2ex}
\caption{Pseudocode for \sys's read wrapper function.
Functions beginning with ``\texttt{raw}'' refer to native (unwrapped) access 
  to the database tables storing the data or the logs.
Identifiers starting with capital letters indicate a member of the log 
structures.}%
\label{fig:alg-read}
\end{figure}

%% file: alg-write.tex
\begin{figure}
\hrule
\begin{lstlisting}[language=iPython]
 def |write|(table, key, val):
     logKey = [$ID$, $STEP$]
     linkedDAAL = |rawScan|(table,
         cond: "Key is {key}"
         project: ["RowId", "NextRow", 
                   "RecentWrites.{logKey}"])
     if logKey not in linkedDAAL:
         tail = |getTail|(linkedDAAL)
         |tryWrite|(table, key, val, tail)
     $STEP$ = $STEP$ + 1
 def |tryWrite|(table, key, val, row):
     logKey = [$ID$, $STEP$]
     ok = |rawCondWrite|(table, row[RowId],
         cond: "({logKey} not in RecentWrites)
                && (LogSize < N)",
         update: "Value = {val};
                  LogSize = LogSize + 1;
                  RecentWrites[{logKey}] = NULL")
     if ok: # Case B
         return
     row = |rawRead|(table, row[RowId])
     if logKey in row[RecentWrites]: # Case A
         return
     elif row[NextRow] does not exist: # Case D
         row = |appendRow|(table, key, row)
     else: # Case C
         row = |rawRead|(table, row[NextRow])
     |tryWrite|(table, key, val, row)
\end{lstlisting}
\hrule\vspace{2ex}
\caption{Pseudocode for \sys's write wrapper function.}%
\label{fig:alg-write}
\end{figure}

%% file: fig-writestate.tex
\begin{figure}[t]
\begin{subfigure}{0.73\columnwidth}
\footnotesize
\begin{tabular}{lccc}
 & logKey $\in$ logs & $\text{logSize} < N$ & $\exists$ nextRow \\
\toprule
A & True & * & * \\
B & False & True & False \\
C & False & False & True \\
D & False & False & False \\
\bottomrule
\end{tabular}
\caption{Cases}
\label{fig:write_states_cases}
\end{subfigure}
\hfill
\begin{subfigure}{0.23\columnwidth}
\includegraphics[width=\columnwidth]{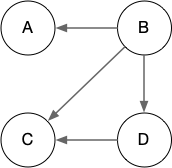}
\caption{Transitions}
\label{fig:write_states_transitions}
\end{subfigure}
\caption{Possible cases for the state of a candidate tail in the linked
DAAL during a \texttt{write} and its potential transitions.}
\label{fig:write_states}
\end{figure}

%% file: alg-syncinvoke.tex
\begin{figure}
\hrule
\begin{lstlisting}[language=iPython,escapeinside={(*}{*)}]
 def |syncInvoke|(callee, input):
     calleeId = |UUID|()
     logKey = [$ID$, $STEP$]
     $STEP$ = $STEP$ + 1
     ok = |rawCondWrite|($InvokeLog$, logKey,
         cond: "{logKey} not in InvokeLog"
         update: "Id = {calleeId};
                 Result = NULL")
     if not ok:
         record = |rawRead|($InvokeLog$, logKey)
         calleeId = record[Id]
         result = record[Result]
     if result does not exist:
         return |rawSyncInvoke|(callee,
             [calleeId, input])
 # When the Callee is done it issues a callback 
 # to the caller. Below is the callback handler.
 def |syncInvokeCallbackHandler|(calleeId, result):
     |rawWrite|($InvokeLog$, cond: "Id = {calleeId}",
              update: "Result = {result}")
\end{lstlisting}
\hrule\vspace{2ex}
\caption{Pseudocode for synchronous invocation of other SSFs.
Asynchronous invocations are similar, but since they do not have return values,
  the callback is invoked as soon as the callee logs the intent in its intent table.
We give the code for the callee's actions in 
  Appendix~\ref{s:appendix-impl}.}%
\label{fig:alg-synccall}
\end{figure}

%% file: fig-callback.tex
\begin{figure}
  \centering
  \vspace{-1em}
  \includegraphics[width=0.41\textwidth]{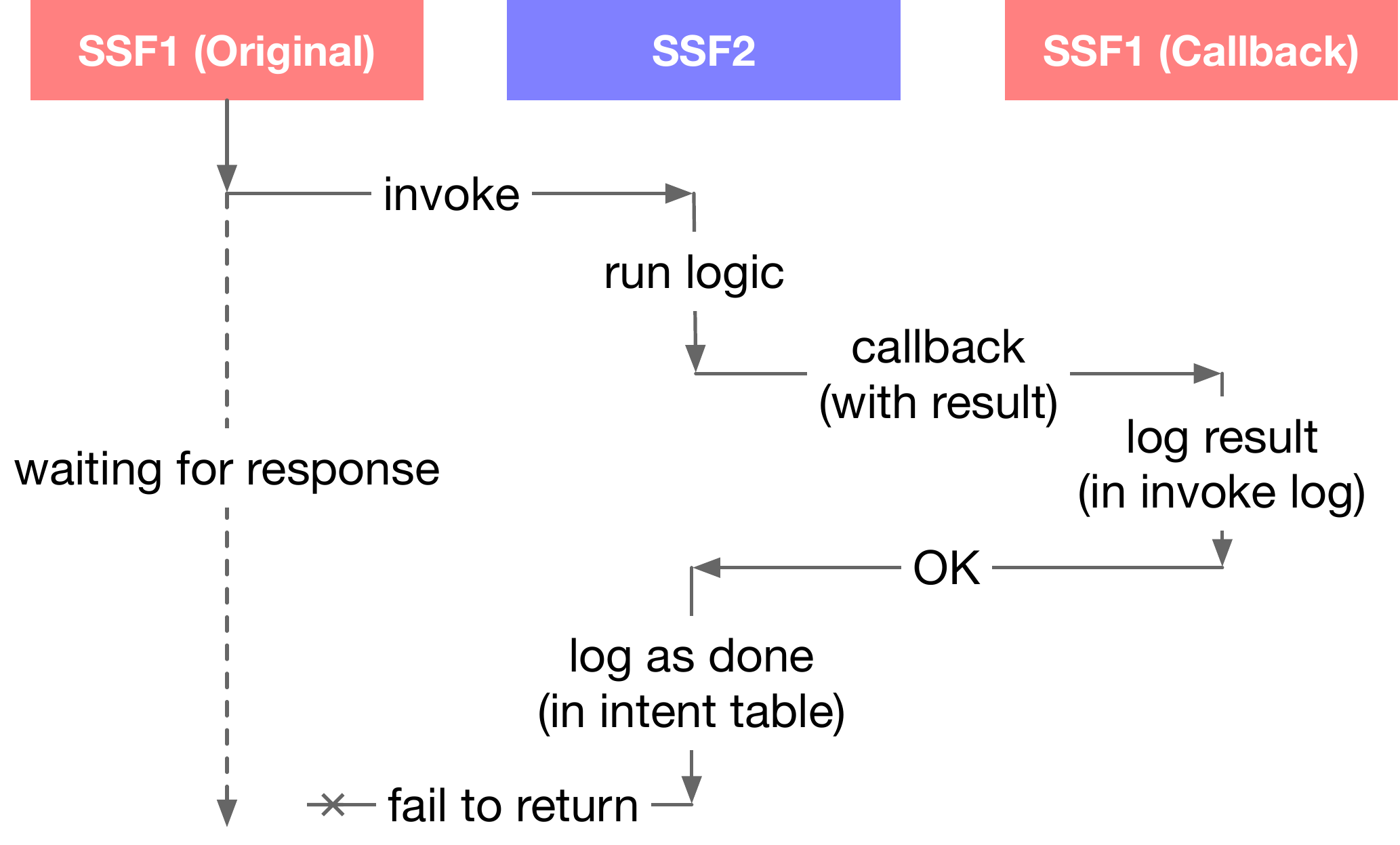}
  \caption{SSF1 synchronously invokes SSF2, which then fails 
  to return after logging the operation as done. 
  The callback ensures that SSF1 has the result of SSF2 
  before SSF2 marks itself as done.}%
\label{fig:callback}
\end{figure}

%% file: collectors.tex
\section{Garbage Collection}\label{s:garbage}

If left alone, the linked DAAL will grow indefinitely.
While \sys's use of scans means that the linked DAAL's length is generally not
  the performance bottleneck, unbounded growth of the linked
  DAAL and logs (intent table, read log, invoke log) can lead 
  to significant overheads and storage costs.
\sys{} ensures that logs are pruned and the linked DAAL remains shallow 
  with a garbage collector (GC) that deletes old rows and log entries
  without blocking SSFs that are concurrently accessing the list.
The GC is an SSF triggered by a timer.

At a high level, the protocol has six parts.
First, the GC finds intents that have finished since the
  last time a GC instance ran and assigns them the current time 
  as a finish timestamp.
Second, the GC looks up all intents whose finish 
  timestamp is `old enough' (we expand on this next), and marks them 
  as `recyclable.'
Third, the GC removes log entries (in the read and invoke logs)
  that belong to recyclable intents.
Fourth, the GC disconnects, for every item, the non-tail rows of
  their linked DAAL that have empty logs, marks
  these rows as `dangling', and assigns them the current time as
  a dangling timestamp.
Fifth, the GC removes all rows whose dangling timestamp is `old enough.'
Finally, the GC removes the log entries from the intent table.
The algorithm is given in Figure~\ref{fig:gc}, with more details in 
  Appendix~\ref{s:appendix-impl}.
Note that GCs only need at-least-once semantics in order to avoid
  memory leaks in the presence of crashes; they do not 
  use \sys's exactly-once API\@.
Instead, GCs defer the removal of entries in the intent table until the end.

\input{alg-gc}

\heading{Assumption}
The safety of garbage collection relies on a synchrony assumption.
In particular, it assumes that an individual SSF instance terminates, one way 
  or another, in at most $T$ time.
This allows the GC to delete the logs of completed intents after
  waiting $T$ time for all running instances of the completed intents to 
  finish.
Note that no new instances will be started by an SSF's IC 
  after the intent is marked as `done.'

Our assumption is based on the observation that serverless providers 
  enforce user-defined execution timeouts on SSF 
  instances (\S\ref{s:bg:serverless}), but otherwise provide no interface 
  for developers to kill or stop running functions.
We can derive a conservative bound for $T$ from these user-defined timeouts.
Note that even if providers refuse to kill SSFs after the timeout,
  we can work around this issue (at high cost) by having the GC change the
  database's permissions or rename tables so that ongoing SSF 
  instances (including stragglers that stick around
  after the intent is done) fail to corrupt the database; 
  instances that start after the change are fine.

\heading{Safety of concurrent access}
With the above assumption, \sys's GC preserves exactly-once 
  semantics without needing to interrupt SSF instances.
First, observe that an intent is marked as recyclable only after \sys{} 
  is sure that no live SSF instance requires the intent.
Accordingly, the read log, invoke log, and intent table entries 
  for the intent will never be accessed again.
For the linked DAAL, the GC only disconnects a row when all of the contained 
  logs are marked as recyclable and it is not the tail.
New traversals of the linked DAAL for read or write operations will not observe 
  the disconnected row (technically the \texttt{rawScan} operation will
  return these disconnected rows, but they will be ignored during 
  the traversal of the local linked DAAL).
Running SSF and GC instances, however, may be in the process of traversing 
  the disconnected row---if \sys{} deleted it immediately, the SSF or GC 
  might become stranded.
To prevent this, \sys{} keeps the disconnected row for an 
  additional $T$ time to ensure that instances with such 
  references terminate successfully.

\heading{Safety of concurrent modifications}
The linked DAAL also supports garbage collection in the presence of
  concurrent appends from SSFs and deletions from other GC instances
  owing to it being a type of non-blocking linked list.
In fact, it is simpler than traditional non-blocking linked
  lists~\cite{harris01pragmatic, zhang13practical, valois95lock}
  because new rows are always appended to the tail,
  and GCs never touch the tail.
The only interesting case is the concurrent disconnection of neighboring
  rows such as $X$ and $Y$ in $A\rightarrow X\rightarrow Y\rightarrow B$.
In this case, the disconnection of $X$ succeeds, but the disconnection
  of $Y$ will not be visible because the updated \texttt{NextRow} 
  pointer in $X$ is no longer part of the linked DAAL\@.
The next GC run disconnects $Y$ permanently.

%% file: alg-gc.tex
\begin{figure}
\hrule
\begin{lstlisting}[language=iPython]
 def |garbageCollection|():
     time = |now|()
     recyclable = []
     for id, intent in $IntentTable$:
         if intent[Done]:
             if FinishTime not in intent:
                 intent[FinishTime] = time
             elif time - intent[FinishTime] > T:
                 recyclable.append(id)
     for id in recyclable:
         |remove| from $ReadLog$
             where "LogKey[Id] == {id}"
         |remove| from $InvokeLog$
             where "LogKey[Id] == {id}"
     for table, key in |getAllDataKeys|():
         rows = |rawScan|(table,
                        cond: "Key == {key}")
         for row in rows:
             for log in row[RecentWrites]:
                 |mark| if log[Id] in recyclable
             if |fullyMarked|(row[RecentWrites])
                     and row[NextRow] exists:
                 |prev|(row)[NextRow] = row[NextRow]
                 if DangleTime not in row:
                    row[DangleTime] = time 
         rows = |rawScan|(table, cond: "Key == {key}
                     && {time} - DangleTime > T")
         for row in rows:
             if row not reachable from head(key)
                 |delete| row
     for id in recyclable:
         |remove| from $IntentTable$
             where "LogKey[Id] == {id}"
\end{lstlisting}
\hrule\vspace{2ex}

\caption{Pseudocode for \sys's lock-free, thread-safe garbage collection algorithm.
$T$ is the maximum lifetime of an SSF instance.}
\label{fig:gc}
\end{figure}

%% file: transactions.tex
\section{Supporting Locks and Transactions}\label{s:transactions}
In addition to exactly-once semantics, \sys{} also provides
  support for locks and transactions with user-generated aborts.

\subsection{Locks}
\label{s:transactions:locks}

\sys's approach to mutual exclusion borrows an abstraction in Olive
  called ``locks with intent'', where locks over data items are owned
  by an intent rather than a specific client.
This means that, if an SSF instance calls \texttt{lock(item)}
  and then crashes, the lock is not lost and held indefinitely;
  rather, the IC will soon restart the instance.
The re-executed instance, upon arriving at the \texttt{lock(item)} call, will
  see that it already acquired the lock and be able to continue with the remaining operations as
  if the original SSF instance had never crashed.

In \sys, the ownership of a lock on a given item is kept alongside the data and
  logs in the ``lock owner'' column of the item's linked DAAL\@.
Lock acquisition and release are logged to the DAAL as writes to the item using
  \sys's \texttt{condWrite} semantics, where the condition is that the lock is
  either owned by the current SSF or has an empty lock-owner column in the DAAL\@.
The exactly-once semantics are needed for cases where an SSF is re-run after
  successfully releasing a lock.

Note that \sys only guarantees exactly-once semantics---it does not absolve the
  developer from writing bug-free code.
Thus, problems like infinite loops within critical sections and deadlock need to be
  handled with higher-level mechanisms (like the one below) if the user wishes to guarantee liveness.

\subsection{Transactions}\label{s:transactions:tx}

\sys{} uses an extension of the locking mechanism of the preceding section 
  to implement transactions within and across SSF boundaries.
\sys{} transactions are based on a variant of 2PL with wait-die
  deadlock prevention and two-phase commit.
Note that the choice of wait-die (rather than something like wound-wait) is
  deliberate as SSF instances generally cannot kill other instances.
To implement this, we need to track the intent-creation time of each SSF.
We do so by adding to the lock-owner column an intent-creation timestamp and
  checking upon lock-acquisition failure whether the existing lock owner is older
  or younger than the current SSF instance; if older, abort, otherwise, 
  try again (see \figref{fig:tx}).

There are three main parts to \sys's transaction-handling protocol: (1) creating
  and forwarding a \textit{transaction context}, (2) executing \sys{} calls inside
  a transaction, and (3) propagating abort/commit signals throughout a
  workflow.
Note that \sys{} does not currently support \texttt{asyncInvoke} in 
  transactions; however, it does support spawning threads 
  that issue \texttt{syncInvoke} operations and are then joined.

\input{alg-tx}

\heading{Transaction contexts}
In \sys{}, transactions are defined with the \texttt{begin\_tx} and
  \texttt{end\_tx} API calls.
\sys{} assumes that both the begin and the end statements are placed
  in the same SSF, but SSFs can invoke other SSFs inside a transaction, 
  so transactions can span across multiple SSFs.
When an SSF calls \texttt{begin\_tx} it creates a new top-level \emph{transaction
  context} which consists of a unique transaction id and a mode (`Execute', `Commit', or
  `Abort').
Contexts start in `Execute' mode.
The SSF instance will also, upon creating a new context, execute the transaction's
  operations in a new thread/goroutine to catch any runtime exceptions.
The matching \texttt{end\_tx} waits for the result and runs either a commit or
  abort protocol depending on the outcome of the contained operations.

Transaction contexts are passed along with any SSF invocations that occur
  inside the transaction.
Thus, whenever a \sys-enabled SSF starts, it first determines whether it is a part
  of an ongoing top-level transaction by checking whether a context was 
  provided as part of the input.
This is necessary even if the SSF never creates a transaction itself.
If the SSF does create a transaction, the \texttt{begin\_tx}/\texttt{end\_tx} 
  statements will be ignored and all operations will be inherited by the top-level 
  transaction context.
\sys{} does not currently support \emph{nested transaction} semantics~\cite{moss81nested} 
  (e.g., a sub-transaction can abort without causing the top-level 
  transaction to abort).
  
\heading{Opacity}
\sys{} chooses \emph{opacity} as the isolation level for transactions.
Opacity~\cite{guerraoui08on} captures strict 
  serializability~\cite{bernstein79formal, papadimitriou79serializability} with the 
  additional requirement that even transactions that abort do not 
  observe inconsistent state.
The rationale is that observing inconsistent state can lead to undefined behavior
  and infinite loops.
For example, if an SSF instance reads inconsistent state that 
  results in division by zero, it may crash.
\sys's IC will restart the SSF instance and deterministically replay the
  (inconsistent) values to ensure exactly-once semantics, re-triggering the crash.
\figref{fig:alg-opacity} gives another example of how OCC~\cite{kung81optimistic},
  which provides serializability but not opacity, leads to infinite loops.
These issues are not present with isolation levels that
  guarantee that all transactions read from a consistent snapshot.

\input{alg-opacity}

\heading{Operation semantics inside a transaction}
If an SSF is in a transactional context, \sys{} modifies the semantics of
  its API based on the mode to ensure ACID semantics.
We have already discussed two operation modifications that 
  occur in `Execute' mode---one to locks in \figref{fig:tx} and another to 
  \texttt{begin\_tx}/\texttt{end\_tx}, which are ignored.
`Execute' mode also causes \sys{} to call \texttt{lock} before 
  every \texttt{read}, \texttt{write}, and \texttt{condWrite} operation, 
  using the transaction id as the lock holder.
In addition to acquiring locks, \sys{} also changes where reads and writes
  look up and record values.
While lock acquisition still goes to the original tables, \sys{} redirects
  written values to a \emph{shadow table} that acts as a local copy of state
  for the transaction.
Like the original table, this shadow table is also stored as a linked DAAL
  and is garbage collected along with the normal DAAL (except the GC
  also deletes the head and tail).
Unlike the original, the shadow table is partitioned by transaction
  id, with \texttt{Key} relegated to a secondary index.
All \texttt{read} operations check the shadow table first before 
  consulting the real table to ensure that transactions read their own writes.
If, before an operation, an SSF fails to acquire a lock and must
  kill itself (due to wait-die), it returns to its caller 
  with an `abort' outcome.

\heading{Propagation of commit or aborts}
Eventually, a \texttt{begin\_tx}/\texttt{end\_tx} code block will 
  reach the \texttt{end\_tx} with an abort/commit decision.
For commit, \sys{} changes the mode of the context to `Commit',
  flushes the final values of the items in the shadow table to
  the real linked DAAL, and releases any held locks.
\sys{} then calls the SSF's callees and passes them the transaction 
  context in Commit mode.
Note that if an SSF instance fails between flushing the shadow table and 
  notifying the callees of the commit decision, \sys's exactly-once
  semantics ensure that once the SSF instance is re-executed, it will
  pick up from where it left off.
For abort, none of the values have been written to the actual table, so
  \sys{} just releases all locks and invokes all callees in `Abort' mode.

When an SSF is invoked with a transaction context that includes
  a Commit mode, \sys{} skips the SSF's logic, and instead performs only the
  aforementioned commit protocol: flushes the final value of 
  the items, releases any held locks, and 
  notifies its own callees by invoking them 
  with the provided transaction context.
An Abort mode similarly skips the SSF's logic, releases all locks,
  and notifies its callees.
This recursive invocation of callees with a Commit or Abort mode mimics 
  the role of a coordinator in two-phase commit.

\heading{Supporting step functions}\label{s:stepfunction}
The previous discussion assumes a \texttt{begin\_tx} and \texttt{end\_tx} in the
  same SSF\@.
To support transactions across SSFs defined in \textit{step functions}, the
  developers must introduce `begin' and `end' SSFs in their workflow
  (we give an example in Appendix~\ref{s:appendix-impl}).
These SSFs create the transaction context and kickstart the commit or abort 
  protocol.
SSFs that fall between the `begin' and `end' SSFs in the workflow
  execute transactionally.
If an SSF aborts it sends `abort' on its outgoing edges in the workflow; 
  an SSF that receives an abort as input skips its operations and 
  propagates the abort message on its outgoing edges.
This continues until the abort message reaches the `end' SSF, 
  which then sets the transaction context mode to Abort and 
  invokes the `begin' SSF\@.
If `end' executes without receiving any abort message, it sets
  the context mode to Commit instead.
This invocation initiates the second phase of 2PC over the 
  transactional subgraph of the workflow.

\heading{Non-transactional SSFs inside transactions}
While an SSF that does not use transactions can be invoked inside 
  a transaction by another SSF (which automatically forces the non-transactional
  SSF to acquire locks before any accesses), app developers must ensure that 
  the non-transactional SSF is \textit{only} used inside transactional contexts.
Otherwise, non-transactional instances may access the database without
  acquiring locks or obeying the wait-die protocol.

%% file: alg-tx.tex
\begin{figure}
\hrule
\begin{lstlisting}[language=iPython,escapeinside={(*}{*)}]
def |lock|(table, key):
  ok = |condWrite|(table, key,
      cond: "LockOwner = NULL
             || LockOwner.id = TXNID",
      update: "LockOwner = [TXNID, START_TIME]")
  if not ok:
    row = |read|(table, key)
    ownerId, ownerTime = row[LockOwner]
    if ownerTime <= $TXNID$:
      |abort|
    else:
      |lock|(table, key)
\end{lstlisting}
\hrule\vspace{2ex}

\caption{Pseudocode for the lock operation with wait-die deadlock prevention
used during the `Execute' mode of a transaction.}%
\label{fig:tx}
\end{figure}

%% file: alg-opacity.tex
\begin{figure}
\hrule
\begin{lstlisting}[language=iPython,escapeinside={(*}{*)}]
|begin_tx|()
x = |read|("x"); y = |read|("y")
while (x != y): 
  // some logic 
  x++
|write|("x", x + 2); |write|("y", y+4)
|end_tx|()
\end{lstlisting}
\hrule\vspace{2ex}

\caption{OCC leads to an infinite loop when two instances of the above transaction, 
  $T_1$ and $T_2$, execute concurrently. Suppose $x = 0, y=1$ initially.
$T_1$ reads $x=0, y=1$, executes the logic, acquires locks on $x$ and $y$, validates the read set, and 
writes $x=3, y=4$. $T_2$ reads $x=3, y =1$ (corresponding to a state after which $T_1$ updated $x$ 
but before it updated $y$), and is stuck in an infinite loop. Even though $T_2$ is destined to abort,
it will never reach the read set validation step.} 
\label{fig:alg-opacity}
\end{figure}

%% file: eval.tex
\section{Evaluation}\label{s:eval}

\sys{} brings forth an array of programmability and fault-tolerance
  benefits, but with these benefits come costs.
In this section we are interested in answering three questions:
\begin{myenumerate2}
  \item What is the cost of maintaining and accessing the 
  linked DAAL, and how does it compare to applicable baselines?
  \item What are the latency and throughput of representative
  applications running on \sys, and how does \sys{} compare to
  existing serverless platforms that provide neither 
  exactly-once semantics nor transactional support?
  \item What effect does \sys's GC have on linked DAAL traversal, 
  and how does it change as we adjust the timeout ($T$)?
\end{myenumerate2}
  
We answer the above questions in the context of the following
  implementation, applications, and experimental setup.

\subsection{Implementation}

We have implemented a prototype of \sys{} for Go applications that
  runs transparently on AWS Lambda and DynamoDB\@.
In total, \sys's implementation consists of 1,823 lines of Go for the
  API library and the intent and garbage collectors.

\input{apps}

\subsection{Experimental setup}\label{s:eval:setup}

We run all of our experiments on AWS Lambda.
We configure lambdas to use 1 GB of memory and set DynamoDB to
  use autoscaling in on-demand mode.
All of the read and scan operations for \sys{} and the baseline use 
  DynamoDB's strong read consistency.
We turn off automatic Lambda restarts and let \sys's intent
  collectors take care of restarting failed Lambdas.
Our garbage and intent collectors are triggered by a timer every
  1 minute, which is the finest resolution supported by AWS\@.
Note that AWS currently has a limit of 1,000 concurrent Lambdas per
  account. 
As we will see in some of our experiments, this limit is often
  the bottleneck in both the baseline and \sys.
Finally, consistent with our deployability requirement~(\S\ref{s:problem}),
  \sys{} uses no servers.

The baseline for our experiments is running our ported applications on AWS 
  Lambda without \sys's library and runtime.
Consequently, these applications will not enjoy exactly-once
  semantics or support transactions: when running on
  the baseline, the travel reservation app outputs inconsistent results,
  and all apps can corrupt state in the presence of crashes.

\subsection{What are the costs of \sys{}’s primitives?}
\label{s:eval:ops}
We start our evaluation with a microbenchmark that
  measures the cost of each of \sys's primitive operations:
  \texttt{read}, \texttt{write}, \texttt{condWrite}, and \texttt{invoke}.
The keys are one byte and the values are 16 bytes.
We measure the median and 99th percentile completion time of 
  the four operations over a period of 10 minutes at very low load (1 req/s).
As baselines, we also measure the completion time (1) without \sys's exactly-once
  guarantees and (2) using a design that logs writes to a separate table using
  cross-table transactions.
Since \sys's database operations depend on the length of the linked DAAL,
  we populate the chosen key's linked DAAL with a conservative value of
  20 rows, which corresponds to the length of the linked DAAL after 30 minutes
  without garbage collection as described in the experiment of Section~\ref{s:eval:gc}.

\input{fig-op}
Figure~\ref{fig:op} shows
  the overhead of \sys's reads/writes compared to those of the
  baseline stem from two sources: scanning the 
  linked DAAL (instead of reading a single row) and logging.
For invoke, the overheads come from our callback mechanism and 
  logging to the invoke log.
Consequently, all of \sys's operations are around 2--4$\times$ more 
  expensive than the baseline.
In contrast, the approach using cross-table transactions
  does not use a DAAL so \texttt{read}s avoid the scan (but not the
  logging), and \texttt{write}s perform an atomic transaction where 
  the value is written to one table and the log entry is added to another.
The cost of this operation is 2--2.5$\times$ higher than \sys's linked DAAL.
Appendix~\ref{s:appendix-eval} describes the same experiment 
  with a more optimistic setting (5 rows in the linked DAAL); the results are similar.

Note that not all existing databases (e.g., Bigtable)
  support cross-table transactions.
Even for those that do, the performance gain that cross-table 
  transactions have on read operations over using a linked DAAL goes 
  away whenever SSFs use transactions because read locks 
  use \texttt{condWrite} which is a cheaper operation on the 
  linked DAAL\@.

\heading{Other costs}
Another consideration beyond performance is the additional 
  storage and network I/O required by \sys{} to 
  maintain and access all logs and linked DAAL metadata.
For our setup above, the 20-row DAAL for the item takes up 
  8~MB of storage.
Counting all logs and metadata, each operation requires storing between 
  20 to 36 bytes in addition to the value.
In terms of the network overhead introduced by the scan and projection
  approach that we use to traverse \sys's linked DAAL, for a 
  20-row DAAL, each scan fetches 2~KB more data than a baseline
  read to a single cell when measured at the network layer.
Compared to the baseline, Beldi induces one extra scan and write for 
  each read operation, at least one scan for an unconditional write (and potentially
  more scans and writes depending on the scenario), and one read
  and two writes for a function invocation. 
In DynamoDB's on-demand mode, each read costs an additional $\$2.5 \times 10^{-7}$,
    whereas writes cost an additional $\$1.25 \times 10^{-6}$. 
In provisioned-capacity mode, costs are lower but depend on the
    specified capacity.

\subsection{How does \sys{} perform on our applications?}\label{s:eval:apps}

In this section, we discuss the results of our large-scale
  experiments for the movie review and travel 
  reservation services; the social networking site has similar results, 
  so we defer its results to Appendix~\ref{s:appendix-eval}.
The workloads that we use are adapted from 
  DeathStarBench~\cite{gan19open,deathstarbench} with a minor modification
  to support our extended travel reservation service: the transactions to 
  reserve a hotel and flight randomly pick a hotel and a flight out of 
  100 choices each following a normal distribution.
Requests contain random content within the expected schema
  and are generated and measured using wrk2~\cite{wrk2}.

We issue load at a constant rate for 5 minutes, starting at 100 req/s
  and increasing in increments of 100 req/s until the system
  is saturated.
For our applications, we achieve saturation at around 800 req/s.
The primary bottleneck in all cases is compute: AWS enforces
  a limit of 1,000 concurrent Lambdas per account (even if 
  the Lambdas are for different functions), and the HTTP Gateway (or some
  internal scheduler) rejects requests in excess of this limit.

\input{fig-media}
\input{fig-travel}

Figures~\ref{fig:media} and~\ref{fig:travel} depict the results.
In all cases (including the social media app), we observe that, until 
  around 400 req/s (34M per day), \sys's median and 99th-percentile
  response time are each around $2\times$ higher than that of the baseline.
At the highest loads that we could test on AWS, \sys{} still 
  achieves the same throughput as the baseline at a slightly higher
  median response time (around 3.3$\times$ for the travel reservation).
At this high load, \sys's 99th-percentile latency is 
  only 20\% higher for the movie review service, and 80\% higher 
  for the transaction-enabled travel site.
We also test a configuration of the travel site that uses \sys{} for
  fault-tolerance but without transactions.
The median latency at saturation for that configuration is 16\% lower
  and the 99th-percentile latency is 20\% lower than \sys{} with transactions.

\subsection{What is the effect of garbage collection?}\label{s:eval:gc}
\input{fig-gc}

Finally, we evaluate the importance of the choice of garbage collector 
  timeout ($T$) on performance.
Note that this is different from the 1-minute timer that triggers
  the GC SSF~(\S\ref{s:eval:setup}).
$T$ is instead proportional to the maximum lifetime of an SSF and
  determines when a GC can remove a row from the Linked DAAL\@.
Thus, this value is important for safety, whereas the trigger only
  determines when the GC runs.
  
Since $T$ is important to ensure exactly-once semantics, we could
  imagine performing a similar actuarial analysis to those
  involved in setting the end-to-end timeouts of reliable failure 
  detectors~\cite{aguilera09no, leners11detecting}.
However, as Figure~\ref{fig:gc-results} shows, the median
  response times for SSFs that access the linked DAAL are only 
  lightly impacted by the choice of $T$, even as we run the system for 30
  minutes at constant load under pessimistic conditions (all SSF instances
  write to the same key).
As a result, we can be relatively conservative about $T$.
To be clear, this is a testament to the heroic efforts of DynamoDB 
  engineers that have optimized its scan, filter, and projection operations.
Nevertheless, we take some slight credit for ensuring that \sys's linked 
  DAAL is compatible with such operators. 

It is worth noting, however, that while $T$ has a minor impact 
  on performance, it does impact storage overhead and I/O, 
  since \texttt{read} and \texttt{write} operations still fetch a 
  projection of the linked DAAL which scales with the number of rows~(\S\ref{s:eval:ops}).

%% file: apps.tex
\heading{Case studies}\label{s:apps}
To evaluate \sys's ability to support interesting applications
  at low cost, we implement three case studies: a social media
  site, a travel reservation system, and a media streaming and review
  service.
We adapt and extend these applications from DeathStarBench~\cite{gan19open,
  deathstarbench}, which is a recent open-source benchmark suite for
  microservices, and port them to a serverless environment (using Go and AWS Lambda).
This port took around 200 person-hours.
Combined, our implementations total 4,730 lines of Go.
We provide details of the corresponding workflows in
  Appendix~\ref{s:appendix-impl}, and give a brief description below.

\weakheading{Movie review service (Cf. IMDB or Rotten Tomatoes)}
Users can create accounts, read reviews, view the plot and cast 
  of movies, and write their own movie reviews and articles.
Our implementation of this app consists of a workflow of 13 SSFs.

\weakheading{Travel reservation (Cf. Expedia)}
Users can create an account, search for hotels and flights,
  sort them by price/distance/rate, find recommendations, and 
  reserve hotel rooms and flights.
The workflow consists of 10 SSFs, and includes a
  cross-SSF transaction to ensure that when a user reserves
  a hotel and a flight, the reservation goes through only if both
  SSFs succeed.
Note that we extend this app to support flight reservations, as the 
  original implementation~\cite{deathstarbench} only supports hotels.

\weakheading{Social media site (Cf. Twitter)}
Users can log in/out, see their timeline,
  search for other users, and follow/unfollow others.
Users can also create posts that tag other users, 
  attach media, and link URLs.
The workflow consists of 13 SSFs that perform tasks
  like constructing the user's timeline, shortening URLs, handling 
  user mentions, and composing posts.

%% file: fig-op.tex
\begin{figure}
	\centering
  \includegraphics[width=0.45\textwidth]{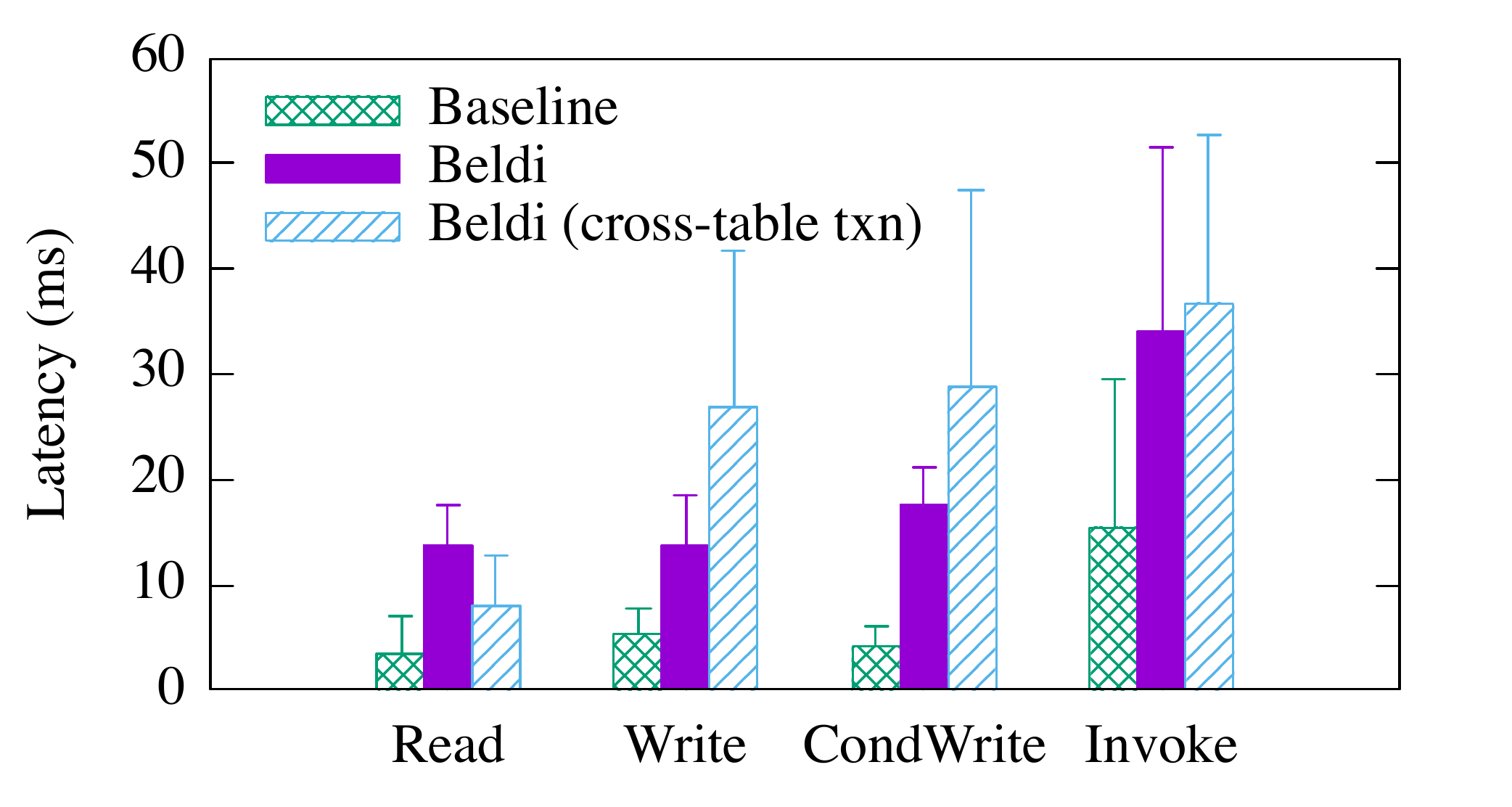}
  \caption{Median latency of \sys's operations. Error bar 
  represents the 99th percentile, and ``cross-table tx'' is an 
  implementation of \sys{} that uses cross-table transactions 
  instead of the linked DAAL.}%
  \label{fig:op}
\end{figure}

%% file: fig-media.tex
\begin{figure}
	\centering
  \includegraphics[width=0.46\textwidth]{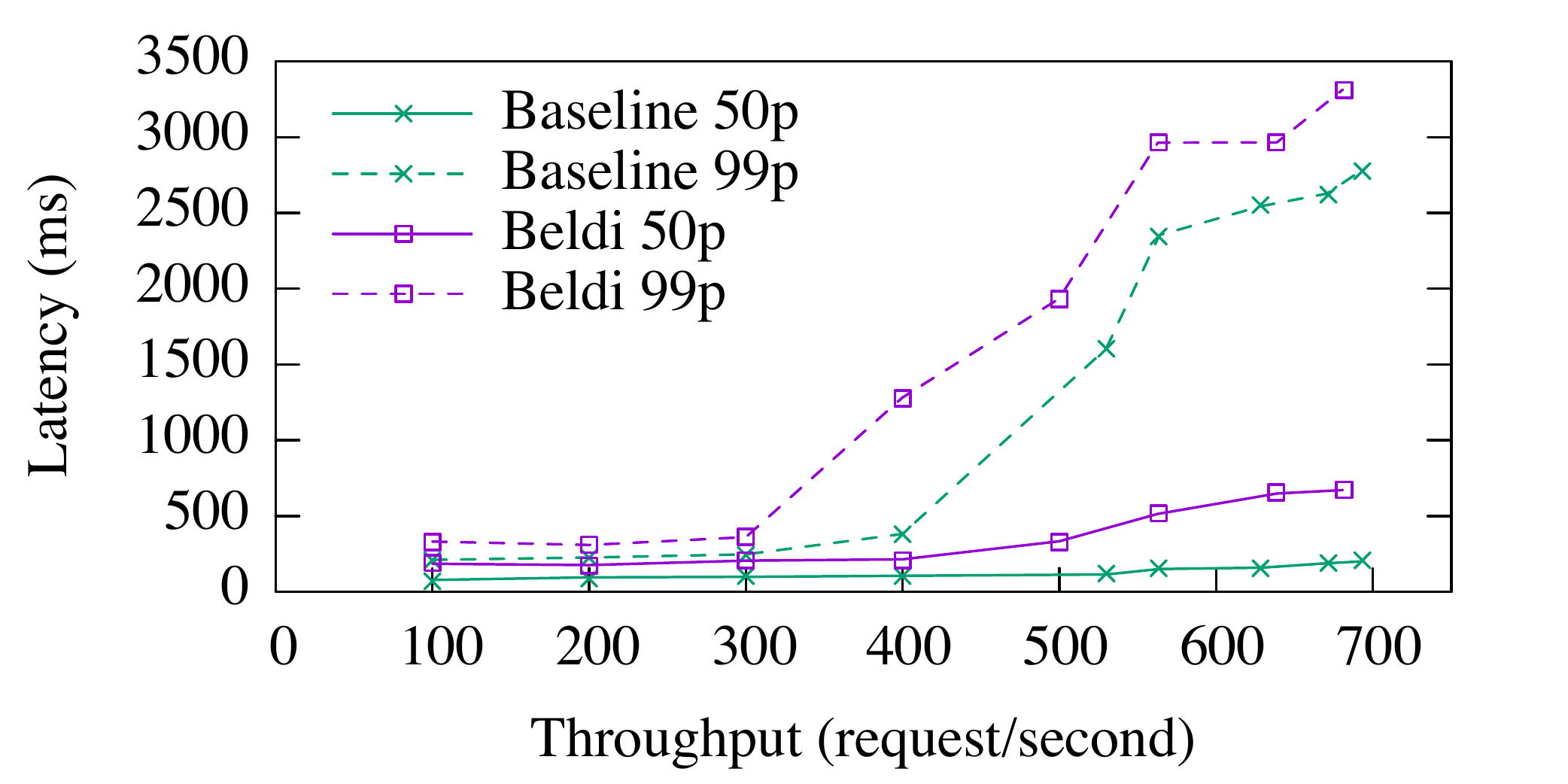}
  \caption{Median response time and throughput for our movie review service.
  Dashed lines represent 99th-percentile response time.}%
  \label{fig:media}
\end{figure}

%% file: fig-travel.tex
\begin{figure}
	\centering
  \includegraphics[width=0.46\textwidth]{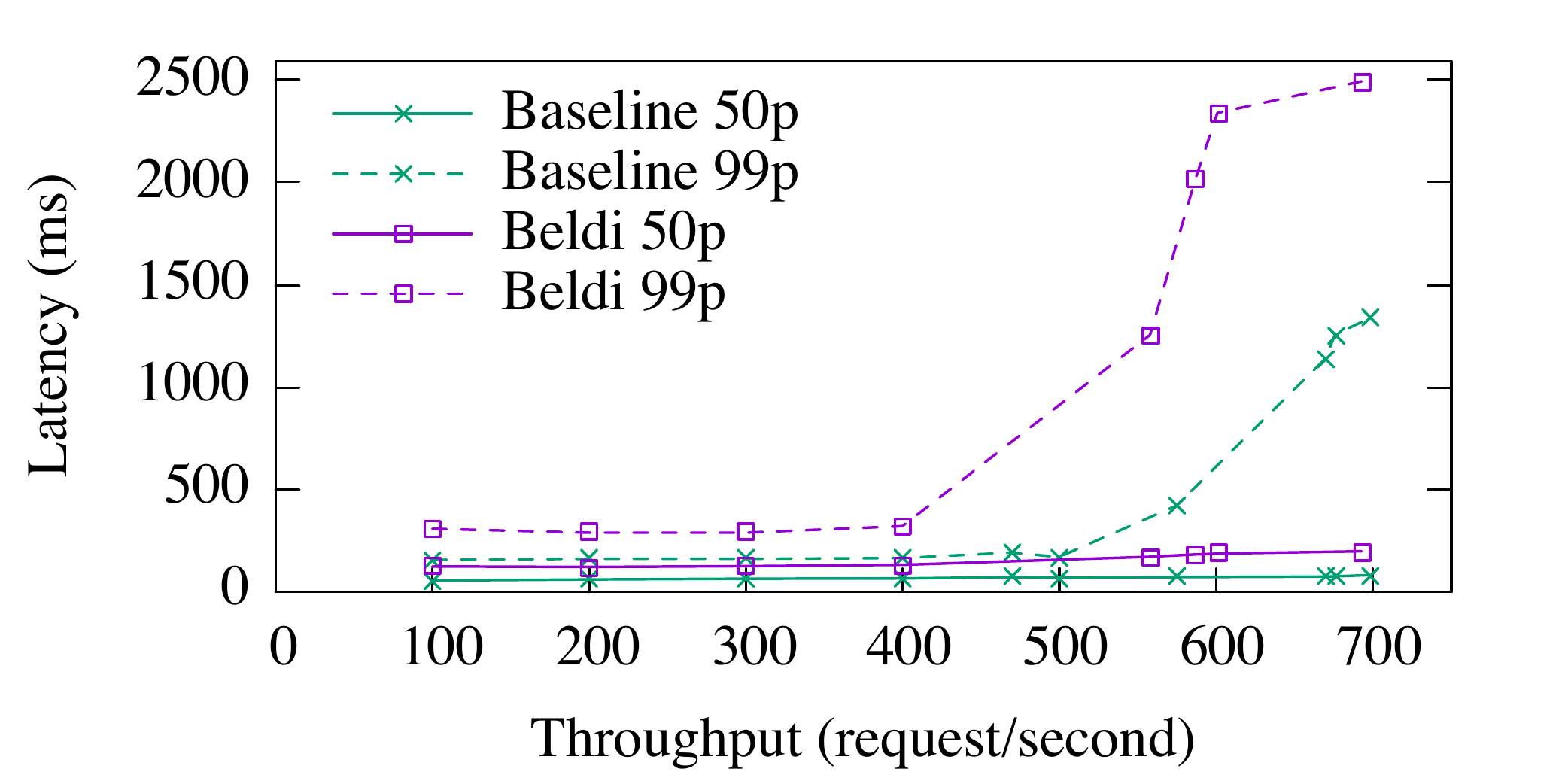}
  \caption{Median response time and throughput for travel reservation service.
  Dashed lines represent 99th-percentile response time.
  \sys{} performs transactions over multiple SSFs to reserve
  a hotel room and a flight, while the baseline returns inconsistent results.}%
  \label{fig:travel}
\end{figure}

%% file: fig-gc.tex
\begin{figure}
  \centering
  \includegraphics[width=0.46\textwidth]{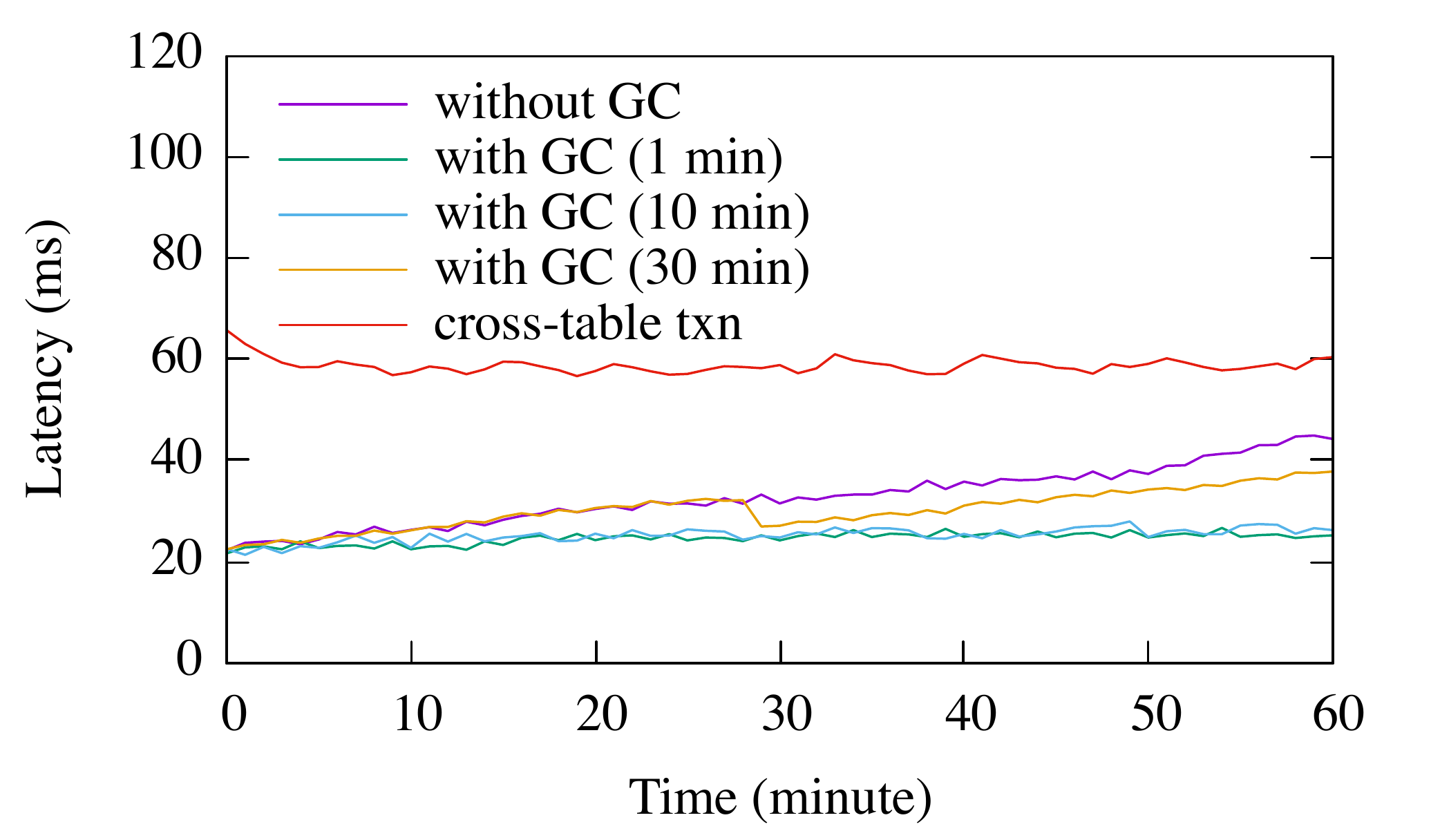}
  \caption{Median response time for an SSF that uses one \texttt{write} operation
  under different GC configurations. Without GC, the linked DAAL grows over time. 
  As a baseline, we configure \sys{} with cross-table transactions 
  that do not use a linked DAAL.}%
  \label{fig:gc-results}
\end{figure}

%% file: discussion.tex
\section{Discussion}

We now discuss a few aspects of \sys, such
as the implication of relying on strongly consistent databases, the
potential benefit of using SQL databases like Amazon Aurora, and the security
implications of SSF federation and reusability.

\heading{Strongly consistent databases}
\sys{} enables developers to write stateful serverless applications 
  without having to worry about concurrency control, fault tolerance, 
  or manually making all of their functions idempotent.
In doing so, \sys{} leverages one or more fault-tolerant databases
  configured to be strongly consistent.
If these databases were to become unavailable, for example due to 
  network partitions, SSFs that write to these unavailable
  databases would also become unavailable until
  the partition was resolved.

\heading{ACID databases}
A natural question is whether SSFs that use ACID databases
  need all of \sys.
For such SSFs, the benefit is not having to maintain a 
  read or write log (or a linked DAAL) since the database does 
  its own logging.
However, ACID databases are not enough to guarantee exactly-once
  semantics for function invocations since they provide 
  atomicity for read and write operations, but have no 
  support for invocations.
As a result, \sys{} would still need to implement mechanisms
  such as callbacks~(\S\ref{s:invoke}) to ensure that a failed 
  SSF is not mistakenly re-executed despite independent 
  garbage collectors.
Furthermore, workflows that contain transactions \emph{across} SSFs
  would still need a collaborative coordination protocol such 
  as the one proposed in Section~\ref{s:transactions:tx}.

\vspace{1ex}\noindent\textbf{Independence of separate applications}
We view SSFs as owning all the data on which they operate, 
  similar to microservice architectures~\cite{microservice-ebook}.
SSFs can isolate the state of different applications by storing 
  each application's state on a different database.
To ensure that a malicious request from one application cannot observe
  the state of another, standard authentication mechanisms
  such as capabilities and public key encryption could be used.

%% file: relwork.tex
\section{Related Work}\label{s:relwork}

We already discuss \sys's differences with Olive~\cite{setty16realizing} throughout.
To summarize, \sys{} builds upon Olive's elegant approach to fault 
  tolerance and mutual exclusion, and adapts it to an entirely new domain.
This adaptation is nontrivial and requires us to introduce
  new data structures, algorithms, and abstractions (e.g., transactions 
  across SSFs).
The result of our innovations is a simple API that SSF developers can
  use to build exciting applications without worrying about fault tolerance, 
  concurrency control, or managing any infrastructure!

In the context of serverless, the observation that existing designs are 
  currently a poor fit for applications that require state has been 
  the subject of much prior work~\cite{hellerstein19serverless,
  klimovic18pocket, wang18peeking, fouladi19from, jangda19formal}.
For example, Cloudburst~\cite{sreekanti20cloudburst} proposes a new architecture
  for incorporating state into serverless functions, and 
  \texttt{gg}~\cite{fouladi19from} proposes workarounds to state-management
  issues that arise in desktop workloads that are outsourced to thousands of 
  serverless functions.
However, the general approach to fault-tolerance in these works is to
  re-execute the entire workflow when there is a crash or timeout---violating 
  exactly-once semantics if any SSF in the workflow is not idempotent.

AFT~\cite{sreekanti20fault} is the closest proposal to \sys{} and 
  introduces a fault-tolerant shim layer for SSFs.
However, AFT's deployment setting, guarantees, and mechanisms are very different.
First, \sys{} runs entirely on serverless functions, whereas AFT 
  requires servers to interpose and coordinate all database accesses.
As a result, \sys{} can run on any existing serverless platform (or even
  in a multi-provider setup) without requiring any modification on their 
  part and without the user needing to administer their own VMs.
Second, \sys{} seamlessly enables transactions within SSFs and across workflows
  with opacity, whereas AFT targets the much weaker (but more efficient) read
  atomic isolation level~\cite{bailis14scalable}.
Due to the weaker isolation, it would be more difficult to implement our 
  travel reservation system on AFT\@.
Finally, \sys{} allows SSFs to be managed independently and to keep 
  their data private from each other, while AFT's servers manage all SSF data, 
  handle failures and garbage collection, and serve as a central point of
  coordination for transactions.

%% file: conclusion.tex
\section{Conclusion}
\sys{} makes it possible for developers to build transactional and 
  fault-tolerant workflows of SSFs on existing serverless platforms.
To do so, \sys{} introduces novel refinements to an existing log-based
  approach to fault tolerance, including a new data structure and algorithms
  that operate on this data structure~(\S\ref{s:linked-daal}), 
  support for invocations of other SSFs with a novel callback 
  mechanism~(\S\ref{s:invoke}), and a collaborative distributed transaction
  protocol~(\S\ref{s:transactions}).
With these refinements, \sys{} extracts the fault tolerance already available
  in today's NoSQL databases, and extends it to workflows of SSFs 
  at low cost with minimal effort from application developers.

%% file: acks.tex
\subsection*{Acknowledgments}

We thank the OSDI reviewers for their feedback and our shepherd, Jay Lorch, 
for going above and beyond and providing suggestions that dramatically
improved the content and presentation of our work. We also thank Srinath Setty for 
many invaluable discussions and his help with Olive. This work was funded in part 
by VMWare, NSF grants CNS-1845749 and CCF-1910565, and DARPA contract HR0011-17-C0047.

%% file: appendix.tex
\section{Additional \sys Implementation Details}
\label{s:appendix-impl}

\input{alg-condwrite}
\input{fig-cwritestate}

\topheading{Conditional Writes}
\figref{fig:alg-condwrite} shows the pseudocode for conditional writes.
Like unconditional writes, this algorithm is built using a transition graph,
  shown in \figref{fig:cwrite_states}, and results in a similar lock-free 
  algorithm.

First, to handle case B$_1$, \sys attempts to write the value to the current
row, gated on the user-defined condition as well as the restriction on current
log length.  If it succeeds, we know that we were in case B$_1$, otherwise we
attempt to log the failure as a false user condition.  A failure of this second
command can only happen if the command has already been executed (case A) or
there is not sufficient room in the logs of the current row (cases C and D).
Also note that, while it is possible for a concurrent SSF to change the value
such that we transition to B$_1$ between the first and second raw conditional
writes, the serialization point is the first raw write---it is valid to record
the false conditional from that check.  After removing B$_1$ and B$_2$ from the
transition graph, the remainder of the cases can be checked similarly to the
unconditional write wrapper.

\input{alg-syncinvoke-callee}
\input{alg-asyncinvoke-callee}

\heading{Invocations}
\figref{alg-synccall} shows the core caller logic for synchronous invocations.
Equally important to correctness is the callee's operation, which is sketched in
\figref{fig:alg-synccallee}.  The SSF instance proceeds normally until the
callee has finished with its execution.  Before the completion is logged to the
intent table, the callee makes a synchronous callback to ensure the result is
properly written to storage in the caller.

In the asynchronous version, the caller's initial \texttt{rawSyncInvoke} is
replaced with a synchronous call to a simple SSF instance that ensures the
intent for the asynchronous call is logged correctly (see
\figref{fig:alg-asynccallee}).  Once the caller is sure that the intent is
logged, it can execute the actual \texttt{rawAsyncInvoke}.  Note that the callee
should not re-execute a completed intent so that the GC can prune the intent
without interference.

\heading{Bounding Intent and Garbage Collection}
Because all executions in \sys, including the collectors, are done in SSFs, the
same execution timeouts that control SSF durations can affect the operation of
these components.  In the worst case, if computing the finish times of the
IntentTable occupies all of the time of the garbage collector, no logs will ever
be pruned.  To address this, we can bound the number of records processed in
each step of both of these functions.  When scanning tables, this can be done
with limits and results paging (e.g., the LastEvaluatedKey in DynamoDB).  When
traversing the DAAL, a similar mechanism can be constructed.  If a collector has
not yet reached the end of the pages, the next execution will continue where the
last left off.

\input{fig-step}

\heading{Transactions on step functions}
\figref{step} shows the step-function workflow for the transaction
  in \figref{tx}.
The begin and end nodes are SSFs that the developer places at the begin
  and end of each transaction.
All SSFs between these two calls will have the same transaction context.

\section{Workflows of Case Studies}\label{s:appendix-workflows}
Figures~\ref{fig:travel-app}--\ref{fig:social-app} depict the workflows
  of our case study applications.
We outline the description of these applications in \S\ref{s:apps}.

\input{fig-travel-app}
\input{fig-media-app}
\input{fig-social-app}

\section{Additional Evaluation Results}\label{s:appendix-eval}

\begin{figure}
	\centering
  \includegraphics[width=0.48\textwidth]{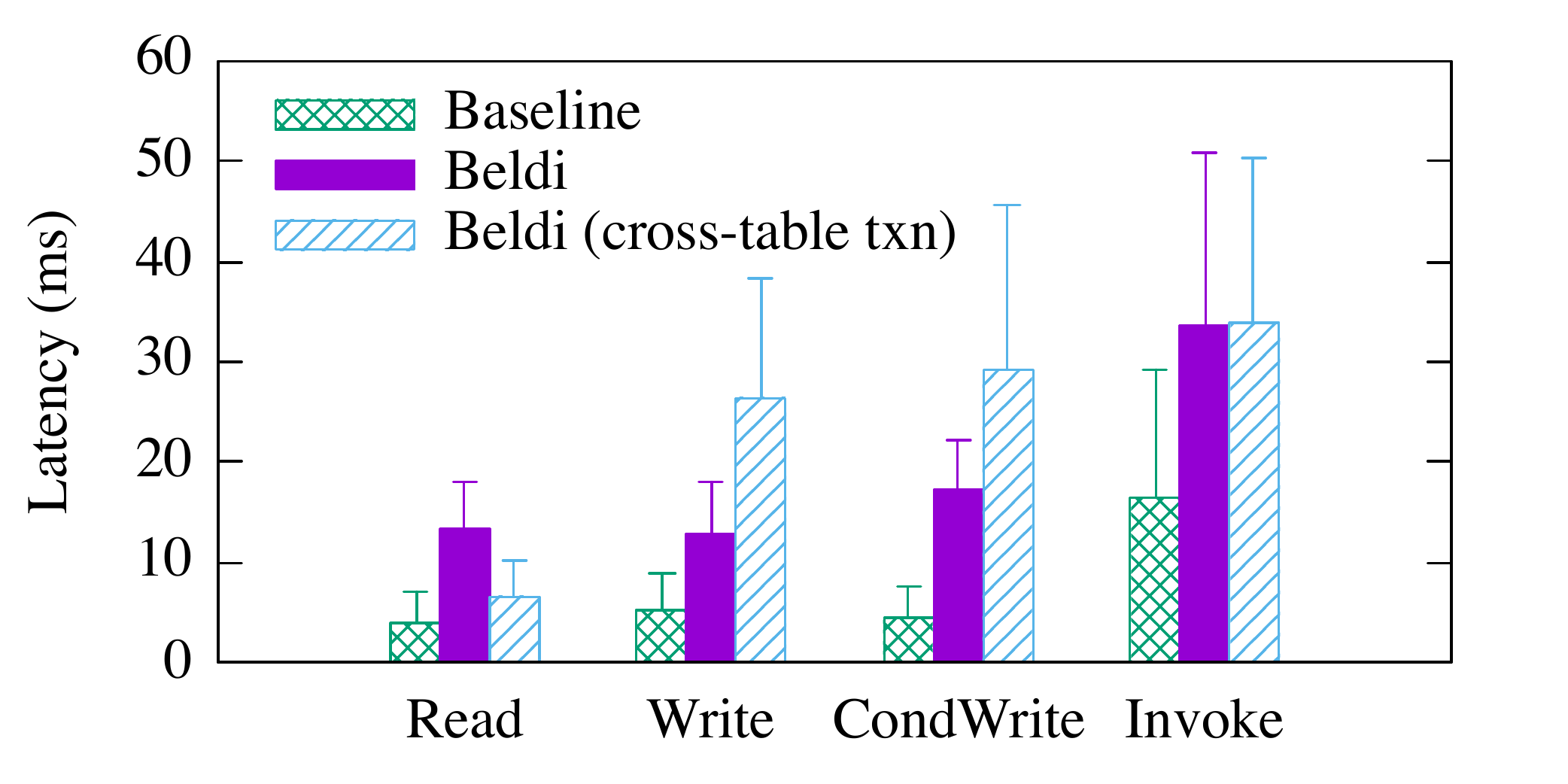}
  \caption{Median latency of \sys's operations. Error bar 
  represents the 99th percentile, and ``cross-table tx'' is an 
  implementation of \sys{} that uses cross-table transactions 
  instead of the linked DAAL.}%
  \label{fig:op5}
\end{figure}

\figref{fig:op5} shows the median and 99th-percentile completion time of the
different operations supported by \sys{} when the linked DAAL has 5 rows.

\subsection{Results for Social Media App}
\input{fig-social}

\figref{fig:social} shows additional results for the latency versus throughput of the social media site.
The results and relative performance of \sys in this experiment are similar to those 
found in Section~\ref{s:eval:apps}.

%% file: alg-condwrite.tex
\begin{figure}
\hrule
\begin{lstlisting}[language=iPython,escapeinside={(*}{*)}]
 def |tryCondWrite|(table, key, val, row, cond):
     logKey = [$ID$, $STEP$]
     ok = |rawCondWrite|(table, row[RowId],
         cond: cond "&& ({logKey} not in RecentWrites)
                     && (LogSize <= N)",
         update: "Value = {val};
                  LogSize = LogSize + 1;
                  RecentWrites[{logKey}] = True")
         )
     if ok: # Case B(*\color{ipython_cyan}$_1$*)
         return True
     ok = |rawCondWrite|(table, row[RowId],
         cond: "({logKey} not in RecentWrites)
                && (LogSize <= N)",
         update: "RecentWrites[{logKey}] = False")
     if ok: # Case B(*\color{ipython_cyan}$_2$*)
         return False
     row = |rawRead|(table, row[RowId])
     if logKey in row[RecentWrites]: # Case A
         return row[RecentWrites][logKey]
     elif row[NextRow] does not exist: # Case D
         row = |appendRow|(table, key, row)
     else: # Case C
         row = |rawRead|(table, row[NextRow])
     return |tryCondWrite|(table, key, val,
                         row, cond)
\end{lstlisting}
\hrule\vspace{2ex}
\caption{Pseudocode for conditional writes.  The \texttt{condWrite} function that calls this one is similar to \texttt{write}.}
\label{fig:alg-condwrite}
\end{figure}

%% file: fig-cwritestate.tex
\begin{figure}[t]
\begin{subfigure}{0.68\columnwidth}
\footnotesize
\setlength{\tabcolsep}{2pt}
\begin{tabular}{lcccc}
 & logKey $\in$ logs & $\text{logSize} < N$ & $\exists$ nextRow & Cond \\
\toprule
A & True & * & * & * \\
B$_1$ & False & True & False & True \\
B$_2$ & False & True & False & False \\
C & False & False & True & * \\
D & False & False & False & * \\
\bottomrule
\end{tabular}
\caption{Cases}
\end{subfigure}
\hfill
\begin{subfigure}{0.3\columnwidth}
\includegraphics[width=\columnwidth]{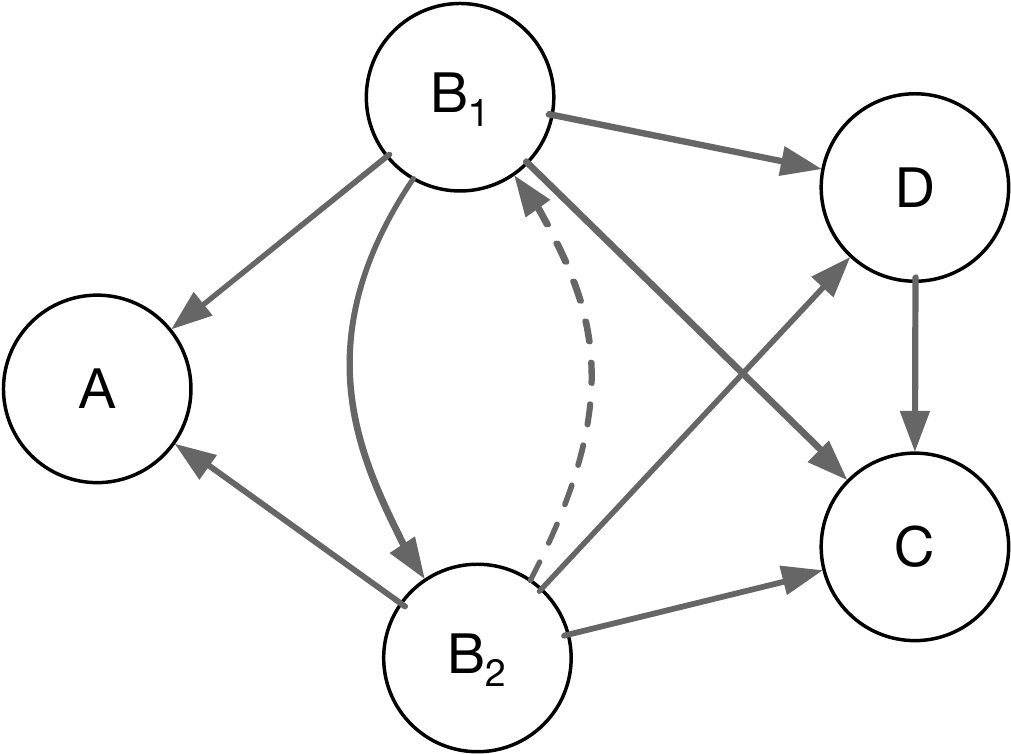}
\caption{Transitions}
\end{subfigure}
\caption{Possible cases for the state of a candidate tail in the linked
DAAL during a \texttt{condWrite} and its potential transitions.}%
\label{fig:cwrite_states}
\end{figure}

%% file: alg-syncinvoke-callee.tex
\begin{figure}
\hrule
\begin{lstlisting}[language=iPython,escapeinside={(*}{*)}]
 def |syncCalleeStub|(caller, calleeId, input):
     $ID$ = calleeId
     $STEP$ = 0
     |log| intent to $IntentTable$
     result = |runMain|(input)
     |rawSyncInvoke|(caller,
         [SYNC_CALLBACK, calleeId, result])
     |log| intent complete to $IntentTable$
\end{lstlisting}
\hrule\vspace{2ex}

\caption{Pseudocode for the callee of a synchronous invocation.}%
\label{fig:alg-synccallee}
\end{figure}

%% file: alg-asyncinvoke-callee.tex
\begin{figure}
\hrule
\begin{lstlisting}[language=iPython,escapeinside={(*}{*)}]
 def |asyncCalleeRegistration|(caller, calleeId):
     |log| intent to $IntentTable$
     |rawSyncInvoke|(caller,
         [ASYNC_CALLBACK, calleeId])
 def |asyncCalleeStub|(caller, calleeId, input):
     $ID$ = calleeId
     $STEP$ = 0
     if calleeId not in $IntentTable$
             or calleeId is complete:
         return
     |runMain|(input)
     |log| intent complete to $IntentTable$
\end{lstlisting}
\hrule\vspace{2ex}

\caption{Pseudocode for the callee of an asynchronous invocation. The first
function is executed via a rawSyncInvoke on the caller.  The second is executed
via a rawAsyncInvoke on the caller after the registration of the intent is
confirmed.}%
\label{fig:alg-asynccallee}
\end{figure}

%% file: fig-step.tex
\begin{figure}
	\centering
  \includegraphics[width=0.4\textwidth]{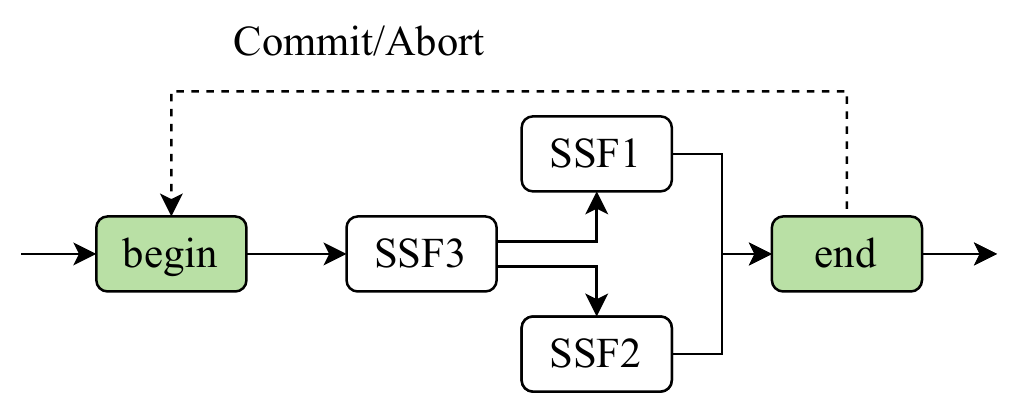}
  \caption{Workflow for the SSFs in \figref{tx} where the
  transaction is defined in the step function. Begin and end are SSFs too.}%
  \label{fig:step}
\end{figure}

%% file: fig-travel-app.tex
\begin{figure}[ht]
	\centering
  \includegraphics[width=0.48\textwidth]{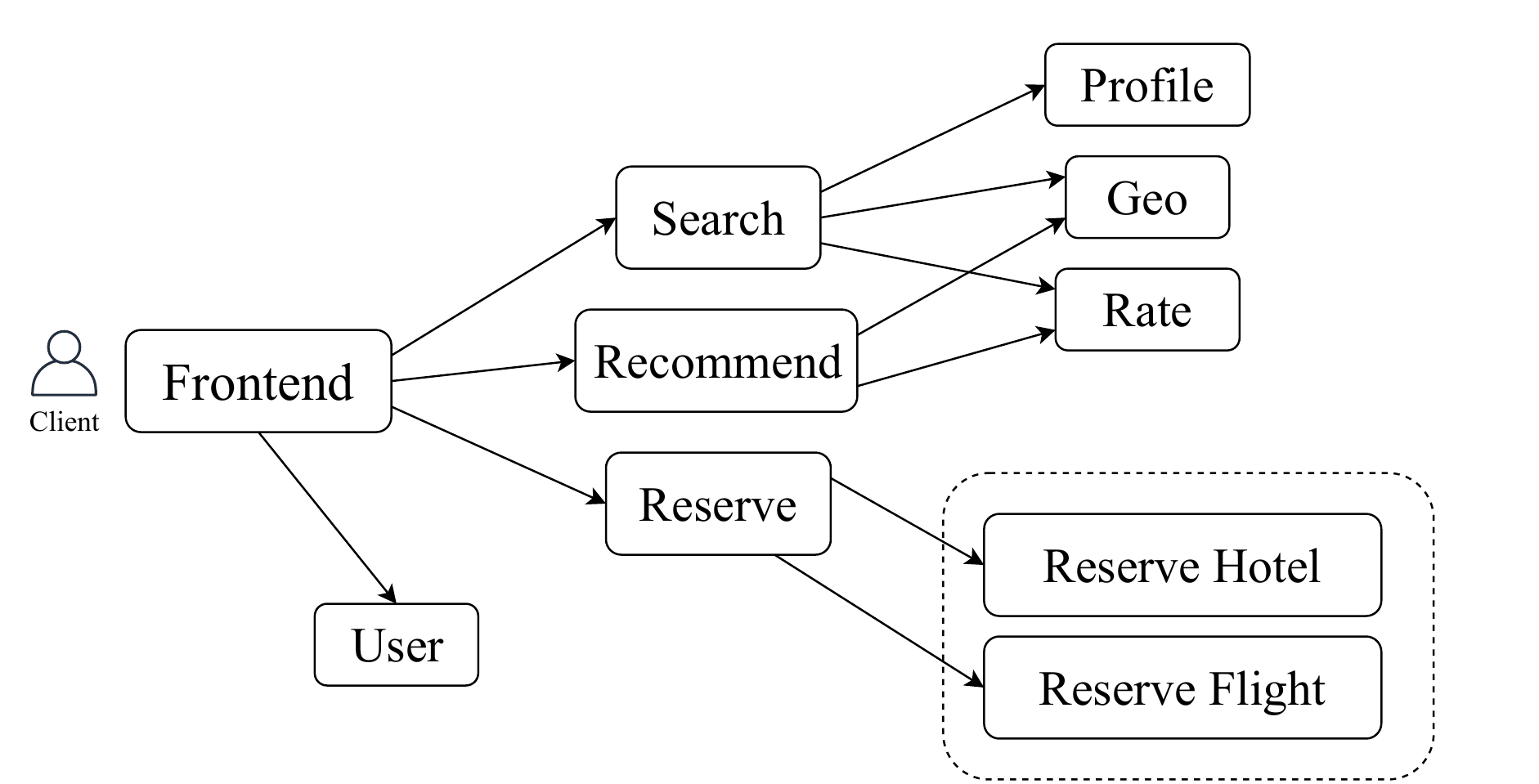}
  \caption{Workflow for travel site app. Reserve Hotel and Reserve Flight are
  invoked within a transaction.}%
  \label{fig:travel-app}
\end{figure}

%% file: fig-media-app.tex
\begin{figure}[ht]
	\centering
  \includegraphics[width=0.48\textwidth]{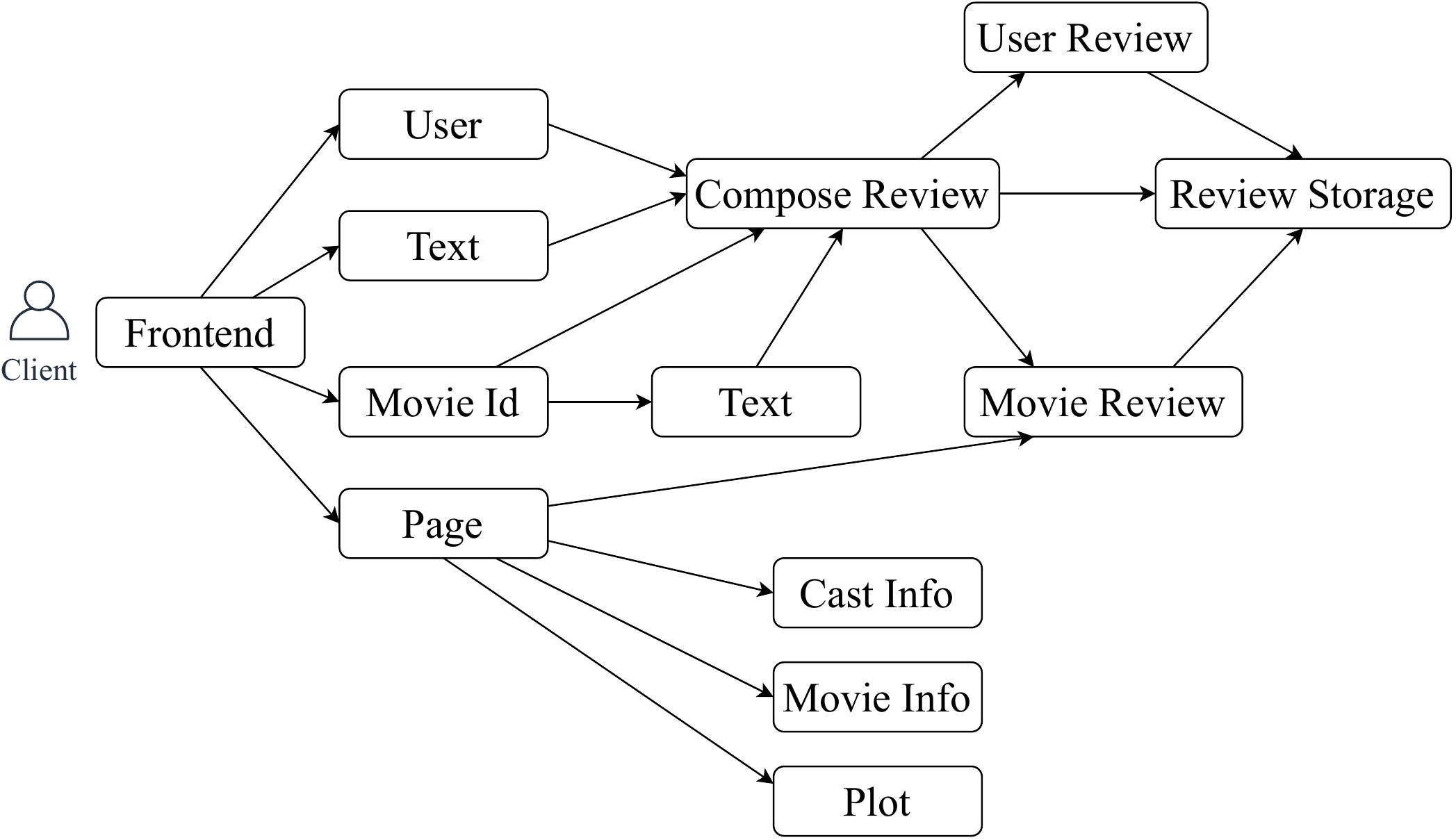}
  \caption{Workflow for media review application.}%
  \label{fig:media-app}
\end{figure}

%% file: fig-social-app.tex
\begin{figure}[ht]
	\centering
  \includegraphics[width=0.48\textwidth]{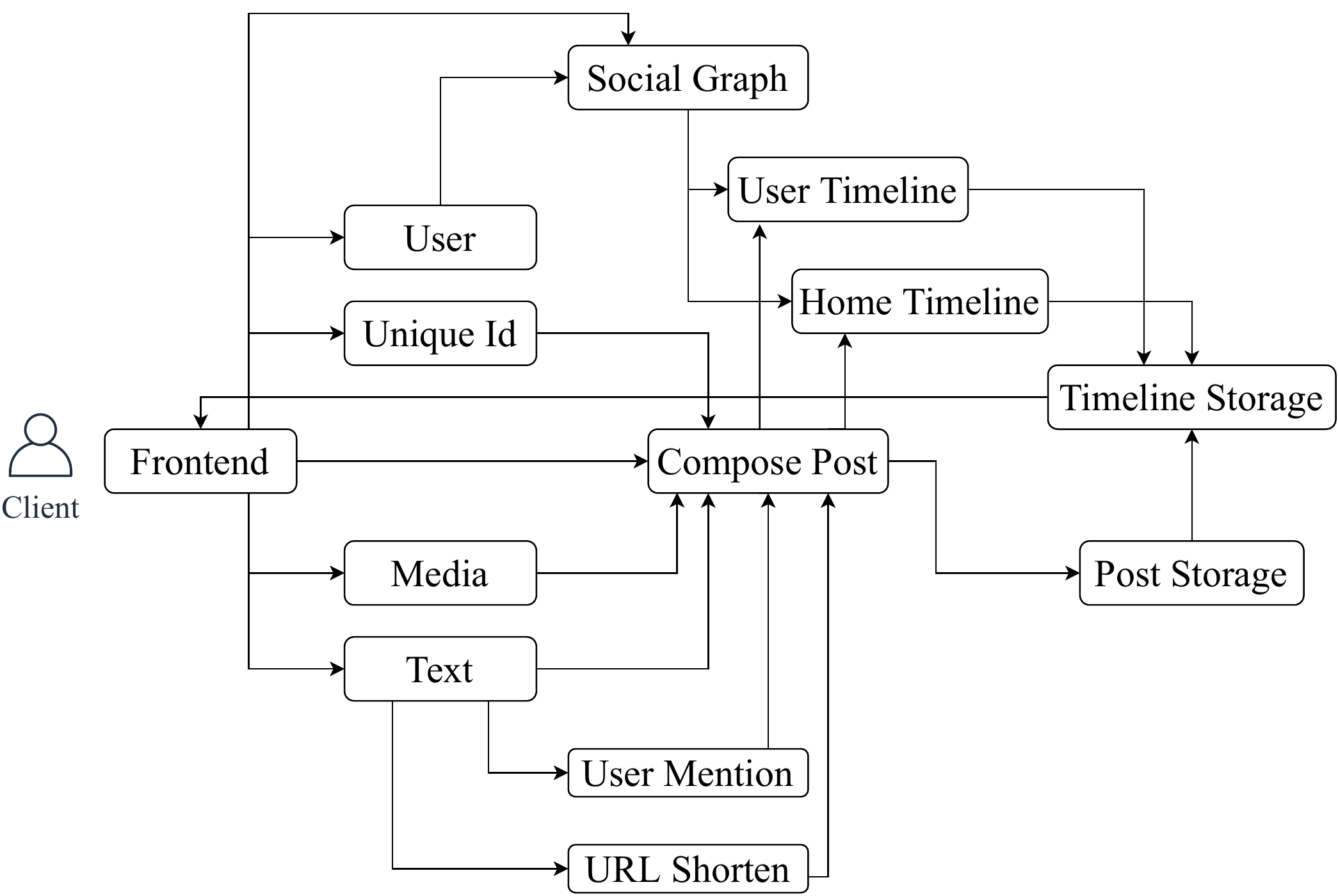}
  \caption{Workflow for social media app.}%
  \label{fig:social-app}
\end{figure}

%% file: fig-social.tex
\begin{figure}
	\centering
  \includegraphics[width=0.48\textwidth]{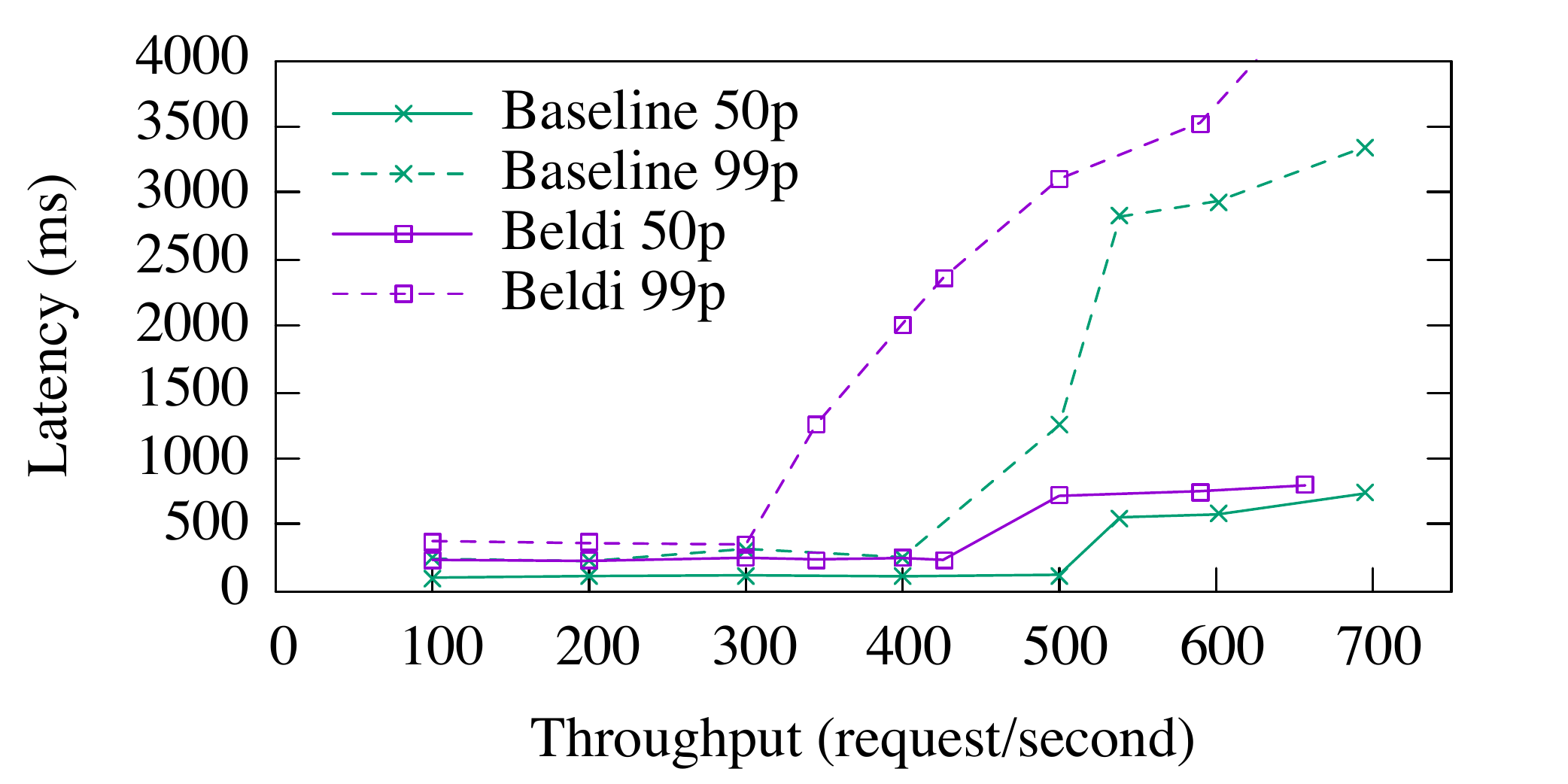}
  \caption{Median response time and throughput for the social media site.
  Dashed lines represent 99th-percentile response time.}%
  \label{fig:social}
\end{figure}